\documentclass[12pt]{article}
\usepackage[dvipsnames]{xcolor}
\usepackage{jcapmod}
\usepackage[utf8]{inputenc}
\usepackage{url}
\usepackage{amsmath}
\usepackage{tikz}
\usepackage{amsfonts}
\usepackage{amssymb} 
\usepackage{bm}
\usepackage{graphicx}
\usepackage{array}
\usepackage{lipsum}
\usepackage{float}
\usepackage{multirow}
\usepackage{hhline}
\usepackage{tabularx}
\usepackage{subfig}
\usepackage{graphicx}
\usepackage{afterpage}
\usepackage{epstopdf}
\usepackage{MnSymbol}
\usepackage[utf8]{inputenc}
\usepackage{mathrsfs}
\usepackage{cleveref}
\usepackage{caption}

\usepackage{amsthm}

\usetikzlibrary{arrows, shapes}
\usetikzlibrary{positioning}
\usetikzlibrary{arrows}
\usetikzlibrary{shapes.multipart}
\newlength{\mytw}
\newtheorem{definition}{Definition}

\newtheorem{conjecture}{Conjecture}

\definecolor{mathyellow}{HTML}{f7e3c3}
\definecolor{mathblue}{HTML}{d1dbe8}

\definecolor{cornellRed}{HTML}{B31B1B}

\begin{document}
\newcommand{\main}{.}
\begin{titlepage}

\setcounter{page}{1} \baselineskip=15.5pt \thispagestyle{empty}
\setcounter{tocdepth}{2}

\bigskip\

\vspace{1cm}
\begin{center}
{\fontsize{22}{20} \bfseries Non-Holomorphic Cycles and \\
\vspace{0.1cm}
Non-BPS Black Branes}

\end{center}

\vspace{0.45cm}

\begin{center}
\scalebox{0.95}[0.95]{{\fontsize{14}{30}\selectfont Cody Long,$^{1,2}$ Artan Sheshmani,$^{2,3,4}$ Cumrun Vafa,$^{1}$ and Shing-Tung Yau$^{2,5}$ \vspace{0.25cm}}}

\end{center}

\begin{center}

\emph{$^1$Jefferson Physical Laboratory, Harvard University \\ Cambridge, MA 02138, USA}\\
\emph{$^2$Center for Mathematical Sciences and Applications, Harvard University \\ Cambridge, MA 02139, USA}\\
\emph{$^3$Institut for Matematik, Aarhus Universitet \\ 8000 Aarhus C, Denmark}\\
\emph{$^4$ National Research University Higher School of Economics, Russian Federation, Laboratory of Mirror Symmetry, \\ Moscow, Russia, 119048}\\
\emph{$^5$Department of Mathematics, Harvard University \\ Cambridge, MA 02138, USA}
\end{center}

\vspace{0.6cm}
\noindent
We study extremal non-BPS black holes and strings arising in M-theory compactifications on Calabi-Yau threefolds, obtained by wrapping M2 branes on non-holomorphic 2-cycles and M5 branes on non-holomorphic 4-cycles. Using the attractor mechanism we compute the black hole mass and black string tension, leading to a conjectural formula for the asymptotic volumes of connected, locally volume-minimizing representatives of non-holomorphic, even-dimensional homology classes in the threefold, without knowledge of an explicit metric. In the case of divisors we find examples where the volume of the representative corresponding to the black string is less than the volume of the minimal piecewise-holomorphic representative, predicting recombination for those homology classes and leading to stable, non-BPS strings. We also compute the central charges of non-BPS strings in F-theory via a near-horizon $AdS_3$ limit in 6d which, upon compactification on a circle, account for the asymptotic entropy of extremal non-supersymmetric 5d black holes (i.e., the asymptotic count of non-holomorphic minimal 2-cycles).

\noindent
\vspace{0.6cm}

\noindent\today

\end{titlepage}
\tableofcontents\newpage

\section{Introduction}

Supersymmetric compactifications of string theory have led to deep insights for both physics and mathematics, as well as the relation between the fields.  These typically involve the choice of a special holonomy manifold, with Calabi-Yau manifolds being a primary example.
Supersymmetric compactifications are the only known examples where there are stable solutions to quantum gravity.  It is thus perhaps not surprising that this is the arena where there are connections to different kinds of mathematical invariants.  For example, wrapping branes of various dimensions on calibrated cycles in such geometries have led to BPS objects, including BPS black holes.   Moreover the count of such black holes (which can also be phrased as mathematical invariants) has led to a microscopic origin of Bekenstein-Hawking black hole entropy, beginning with the work~\cite{Strominger:1996sh}, marking the start of many successful examples of holography in quantum gravity in the framework of string theory.

However, we know that our universe is not supersymmetric at low energies.  Therefore we need to develop tools to study non-supersymmetric configurations in string theory.  This has turned out to be notoriously difficult and there are very few solid results in this direction.  In particular non-supersymmetric configurations are always unstable. Nevertheless, there is a happy middle ground: we can study non-supersymmetric objects in a supersymmetric compactification!  This has the advantage that the theory itself is stable and any instability would be associated to the decay of the non-supersymmetric object in an otherwise stable background. Indeed, the Weak Gravity Conjecture (WGC)~\cite{ArkaniHamed:2006dz}, which postulates that gravity is always the weakest force, is motivated from the assumption that, except for BPS ones, all macroscopic black holes with large enough charge decay to microscopic objects (which may or may not be BPS). The WGC, which is one of the powerful Swampland conjectures, is still at the conjectural level and thus any further studies of non-BPS black holes within string theory would be extremely important for a better understanding of it.  

The mathematical perspective parallels the physical one: the theory of calibrated cycles in manifolds of special holonomy is a long studied subject, with many concrete results.  On the other hand, it is also known that many of the cycle classes in a special holonomy manifold do not admit calibrated representatives.  In such cases one would naturally study minimal volume cycles, which is a notoriously difficult subject in general.  One would expect that at least in the case of special holonomy manifolds the study of non-calibrated minimal cycles should be easier.  Even for such cases, there are very few known results.  It is thus of great mathematical importance if one can improve our understanding for this class. 

A special case of interest is Calabi-Yau 3-folds.  M-theory compactified on such spaces leads to ${\cal N}=2$ supersymmetric theories in 5 dimensions.
Black holes can be constructed by wrapping M2 branes on 2-cycles and black strings can be constructed by wrapping M5 branes on 4-cycles.  It turns out that in these cases known techniques (the ``attractor mechanism"~\cite{Ferrara:1995ih,Ferrara:1996dd,Ferrara:1996um,Strominger:1996kf}) available from study of macroscopic black holes and black strings for large charges (where we rescale the cycle class by a large integer) can give reliable predictions about properties of non-BPS objects and these, in turn, lead to  predictions about existence, stability, and asymptotic count for for minimal volume 2- and 4-cycles in the Calabi-Yau. We now summarize our main results in order to provide the reader with a guide to this work. We attempt to keep this brief and leave the details to the main text.

\subsection{Black branes, minimal cycles, and counting}
We consider non-BPS black holes and strings obtained from compactification of M-theory on a Calabi-Yau threefold $X$. The non-BPS 5d black holes correspond to M2 branes wrapped on non-holomorphic curves in $X$, and the black strings to M5 branes wrapped on non-holomorphic divisors. The main physical tool we use is the black hole/string effective potential, whose minimization determines the values of the moduli at the horizon. We consider cycles for which the minimization procedure fixes the moduli strictly interior to the K\"ahler cone, and so by fixing the asymptotic moduli to be the same as those on the horizon we are able to read off the black hole mass/black string tension, which gives a conjectural formula for the volume of the non-holomorphic cycle wrapped by the brane.  This prediction is expected to be exact in the limit of the asymptotically large charges ($Q\rightarrow N Q$ with $N>>1$).

The wrapped cycle $\Sigma$ is conjectured to be a connected, locally volume-minimizing representative of its homology class $[\Sigma]$. In the case of black holes, in the examples we consider we find that the black holes correspond to local, but not global, volume minimizers of the corresponding curve classes, as there is always a disconnected, piecewise-calibrated representative $\Sigma^\cup$ with smaller volume. The piecewise-calibrated representative (union of holomorphic and anti-holomorphic curves) corresponds to the BPS-anti-BPS constituents of the black hole, and the fact that $\mathrm{vol}(\Sigma^\cup) < \mathrm{vol}(\Sigma)$ shows that the Weak Gravity Conjecture (WGC)~\cite{ArkaniHamed:2006dz} is satisfied in these examples; that is, the non-BPS black holes can decay, and an allowed decay channel is into widely-separated BPS and anti-BPS particles.

In the case of black strings (divisors), we find that in some examples $\mathrm{vol}(\Sigma^\cup) < \mathrm{vol}(\Sigma)$, and so the WGC is satisfied by BPS and anti-BPS states, but in other examples we find $\mathrm{vol}(\Sigma^\cup) > \mathrm{vol}(\Sigma)$ for all piecewise-calibrated $\Sigma^\cup$. Geometrically, the latter case indicates that the holomorphic and anti-holomorphic constituents of the class $[\Sigma]$ fuse to make a smaller cycle; that is, $[\Sigma]$ undergoes ``\textit{recombination}". Here the WGC makes a non-trivial prediction: there should be a stable, non-BPS string in the spectrum. The stable non-BPS string is expected, based on the WGC, to have a small charge. In other words, a non-BPS $\Sigma$ with a large charge will decay to non-BPS strings with smaller charges, which ultimately lead to stable non-BPS low charge string remnant(s).

We also compute the entropy of non-BPS black holes. Morally, this should be related to the asymptotic count of non-holomorphic curves associated to the non-BPS black holes. In addition to the charge (i.e., the homology class of M2 brane) such black holes also carry an $SO(4)$ spin.  To compare with the BPS case, note that in the BPS case, the GV invariants
~\cite{Gopakumar:1998ii,Gopakumar:1998jq} count the $SU(2)_L\subset SO(4)$ spin content of the supersymmetric spinning BPS black holes.
In these cases the asymptotic expansions of the GV invariants are expected to give rise to the semi-classical predictions of  the entropy of supersymmetric spinning BPS black holes
~\cite{Katz:1999xq}.   For simplicity here we limit ourselves to the non-spinning non-BPS extremal black holes.  We expect the extension to spinning case should not be difficult (see e.g.,~\cite{Astefanesei:2006dd} for some progress relating the attractor mechanism to spinning non-BPS black holes).

\subsection{The central charge of non-BPS strings and black hole entropy}

Our second main result is the calculation of the central charge of non-BPS strings and its relation to non-BPS extremal black holes. For elliptic CY 3-folds, we calculate the string central charge in two different ways, that is, using both 5d and 6d supergravity theories, and the results agree. In 6d, obtained by compactifying IIb on an F-theory base $B$, we can construct black strings from D3 branes wrapping a curve $C\subset B$. In both the BPS and non-BPS strings we can obtain the central charge from the attractor values of the moduli, by evaluating the Brown-Henneaux formula~\cite{cmp/1104114999} in the near-horizon $AdS_3$ geometry of the strings. 

\begin{figure}
\centering
\begin{tikzpicture}[sharp corners=2pt,inner sep=7pt,node distance=.8cm,every text node part/.style={align=center}]
\node[draw, rounded corners, minimum height = 2cm, minimum width = 5cm, fill=mathyellow] (state0){
6d BPS Black String};
\node[draw,below=2cm of state0, rounded corners, minimum height = 2cm, minimum width = 5cm, fill=mathyellow](state2){
5d BPS Black Hole};
\node[draw,right=2cm of state0, rounded corners, minimum height = 2cm, minimum width = 5cm, fill=mathblue](state1){
6d Non-BPS Black String};
\node[draw,below=2cm of state1, rounded corners, minimum height = 2cm, minimum width =5cm, fill=mathblue](state3){
5d non-BPS Black Hole};

\draw[-triangle 60] (state0) -- (state2) node [midway, above, left = 0.1cm]{$n\geq 0$};
\draw[-triangle 60] (state1) -- (state3) node [midway, above, right = 0.1cm]{$arbitrary \ n $};
\draw[-triangle 60] (state0) -- (state3) node [midway, above, rotate = -32]{ $n < 0$};
\end{tikzpicture}
\caption{The connections between the various black objects we consider in 6d and 5d. An arrow represents a dimensional reduction and matching of the central charge in the various pictures. A BPS black string in 6d descends to either a BPS or a non-BPS black hole in 5d, depending on the relative orientation of the momentum along the $S^1$. A non-BPS black string in 6d descends to a non-BPS black hole in 5d regardless of the sign of the $S^1$ momentum.}\label{fig:connections1}
\end{figure}
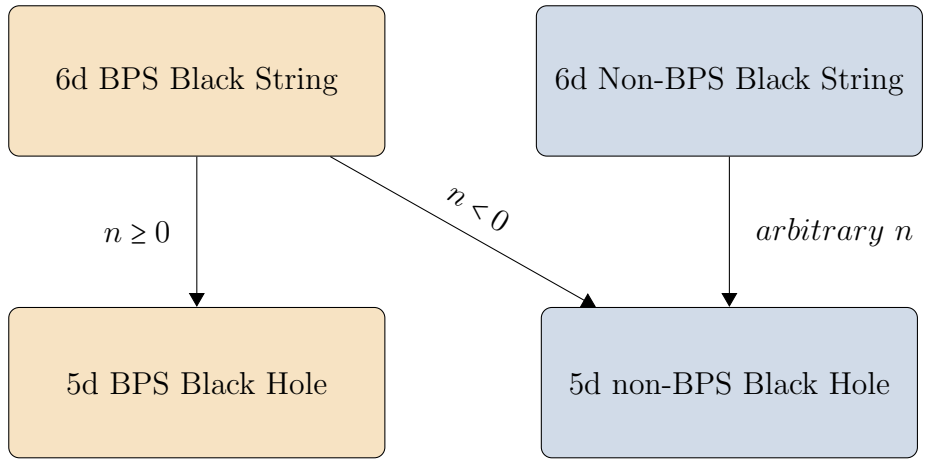
One can then further reduce on an $S^1$ to arrive at a 5d black hole, with $n$ units of momentum along the $S^1$. If one starts from a BPS string in 6d, one can arrive at a BPS or non-BPS black hole depending on the sign of $n$. In both cases this is just a (non)-BPS excitation of a BPS string. However, instead one can start with a non-BPS string in 6d, which then reduces to a non-BPS string in 5d for all $n\neq 0$. We can read off the central charge from the black hole entropy by comparing with the Cardy formula, and in all cases the answer agrees with the central charge computed in the 6d calculation. The relationships between the various black objects are indicated in Fig.~\ref{fig:connections1}.   In both cases the 5d non-BPS black hole entropy is accounted for by the Cardy formula of the 6d string.

The organization for the rest of this paper is as follows:  In \S\ref{sec:tech} we discuss the techniques we use.  In \S\ref{sec:bhexamples} we present examples of black holes for M-theory compactified on CY 3-folds.  In \S\ref{sec:bsexamples} we present examples of black strings.  In \S\ref{sec:f-theory} we show how the non-BPS black holes can arise as BPS and non-BPS black strings in 6d wrapped on a circle and check that this leads to a consistent prediction for its entropy.  In \S\ref{sec:disc} we end the paper with some concluding remarks.

\section{Techniques}\label{sec:tech}
In this section we detail the techniques used to study non-BPS black branes. We first summarize the necessary details of 5d theories obtained from compactification of M-theory on a Calabi-Yau threefold, and introduce the effective potential for black holes and strings. We then relate these black objects to minimal cycles, which is a subject of interest in geometric measure theory. Finally, we propose a physical test for when recombination should occur, by considering the formation of the black brane from its constituent objects. 

\subsection{Black holes and strings in M-theory on a Calabi-Yau threefold}\label{sec:gen}

We consider M-theory compactified on a Calabi-Yau threefold $X$, which yields a 5d effective supergravity theory with eight supercharges~\cite{Cadavid:1995bk}. The field content is organized into a gravity multiplet, vector multiplets, and hypermultiplets, whose scalars parametrize the geometry of $X$. The geometric moduli of a Calabi-Yau threefold are the K\"ahler moduli, counted by the Hodge number $h^{1,1}(X)$, and the complex structure moduli, counted by $h^{2,1}(X)$. The complex structure moduli will not play a role in this work as they decouple from the black brane physics~\cite{deWit:1984rvr,Becker:1995kb} and we will henceforth ignore them, focusing on the K\"ahler moduli which are associated with the vector multiplets. The total number of vectors is $h^{1,1}$, but since one of the vectors belongs to the gravity multiplet, there are only $h^{1,1}(X) - 1$ vector multiplets. The overall volume $\mathcal{V}$ belongs to a hypermultiplet, while the remaining K\"ahler moduli form the real scalar components of the vector multiplets. Together, these $h^{1,1}(X) - 1$ scalars form a \textit{very special geometry}~\cite{deWit:1992cr}, where we can parametrize the moduli space by $h^{1,1}$ real fields $t^I$, subject to the constraint that the overall volume is constant. Omitting the hypermultiplets, the bosonic portion of the effective action takes the form
\begin{equation}\label{eqn:action5}
S_5 = \frac{1}{2\kappa_5^2}\int\left( R\ast\mathbf{1} -G_{IJ}d t^I \wedge \ast d t^J - G_{IJ} F^I \wedge \ast F^J - \frac{1}{3!}C_{IJK}F^I \wedge F^J \wedge A^k\right)\, ,
\end{equation}
where the $t^I$ are the K\"ahler moduli, parameterizing the K\"ahler form as $J = t^I \omega_I$, $\omega_I \in H^{1,1}(X)$, the $C_{IJK}$ are the triple intersection numbers of the divisors $D_I$, dual to the $\omega_I$, and $F^I = dA^I$ are the field strengths for the $U(1)$ gauge fields. The gauge kinetic function and metric on moduli space are written as
\begin{equation}
G_{IJ} = -\frac{1}{2}\partial_I \partial_J \mathrm{log}(\mathcal{V})\, ,
\end{equation}
and the overall volume is written as
\begin{equation}
\mathcal{V} = \frac{1}{6}\int\limits_X J \wedge J \wedge J = \frac{1}{6}C_{IJK}t^I t^J t^K\, .
\end{equation}
To enforce the constraint that $\mathcal{V}$ is constant it is convenient to set $\mathcal{V} = 1$, and so all curve and divisor volumes are measured relative to the overall volume. We will switch between explicitly and implicitly setting $\mathcal{V} = 1$, as some formulas are more easily presented with $\mathcal{V}$ left unfixed. 
Let $D$ denote the divisor dual to the K\"ahler form $J$. The gauge kinetic function can then be expressed as
\begin{equation}
G_{IJ} = \frac{1}{2\mathcal{V}^2} (-A_{IJ} \mathcal{V} + \tau_I \tau_J)\, ,
\end{equation}
where 
\begin{equation}
A_{IJ} = D_I \cdot D_J \cdot D = C_{IJK}t^K\,,
\end{equation} 
which can be thought of as the volume of the intersection of the divisor $D_I$ with the divisor $D_J$, when they intersect transversely, and
\begin{equation}
\tau_I = \frac{1}{2}D_I \cdot D^2 = \frac{1}{2}C_{IJK}t^J t^K\, ,
\end{equation} 
is the volume of the divisor $D_I$. The inverse gauge kinetic function takes the form
\begin{equation}
G^{IJ} = 2\left(-\mathcal{V}A^{IJ} + \frac{t^I t^J}{2}\right)\ ,
\end{equation}
where $A^{IJ}$ is the inverse matrix of $A_{IJ}$. 

The charged objects arise from M2 and M5 branes wrapping cycles of appropriate dimension in $X$. Electrically charged particles arise from M2 branes wrapping curves, or real two-cycles, in $X$, while magnetically charged strings arise from M5 branes wrapping surfaces, or real four-cycles, in $X$. When discussing black holes we will fix units so that the mass of a BPS electrically charged particle is given by the volume of the corresponding curve wrapped by the M2 brane, so the electric central charge takes the form
\begin{equation}
Z_e = q_I t^I\, , \quad I = 1,\dots,h^{1,1}\, ,
\end{equation}
where the $q_I$ are the quantized electric charges, given by the intersection number of the curve with the divisor $D_I$. 
Similarly, when we consider black strings we will fix units for the magnetic central charge so that the tension of a BPS string is given by the volume of the corresponding surface wrapped by the M5 brane, and so the magnetic central charge takes the form
\begin{equation}
Z_m =  p^I \tau_I\, , \quad I = 1,\dots,h^{1,1}\, ,
\end{equation}
where the $p^I$ are the wrapping numbers of the M5 brane around the $I$-th divisor, which reflects the underlying homology class. 

Macroscopic black holes and black strings arise from introducing an appropriately large amount of the respective charges. Extremal black holes, which have the minimum amount of mass given their charge, have been studied extensively in the context of string theory (see e.g.,~\cite{Strominger:1996sh,Vafa:1997gr,Tripathy:2005qp,Larsen:2006xm}). The presence of the large black object back-reacts on the vector multiplet moduli, and forces the moduli to flow to fixed values at the horizon determined only by the charges. This is known as the attractor mechanism. For an attractor these moduli values are independent of those set at asymptotic infinity, and the entropy of the black object is only sensitive to the attractor values themselves. For extremal black holes, the attractor values are set by minimizing the black hole effective potential~\cite{Chou:1997ba,Larsen:2006xm,Ferrara:2006xx}, which can be written as
\begin{equation}
V_{eff} =G^{IJ}q_I q_J\, .
\end{equation}
Physically this can be thought of as minimizing the mass of the black hole with a given set of charges, when the asymptotic moduli are set to the attractor values. We will focus on attractor solutions in this work. The black hole effective potential can also be written as
\begin{equation}
V_{eff} =  \frac{2}{3}Z_e^2+ G^{IJ}(\mathcal{D}_I Z_e)(\mathcal{D}_J Z_e)\, ,
\end{equation}
where
\begin{equation}{}
\mathcal{D}_I Z_e := (\partial_I -\frac{1}{3}\partial_I \mathrm{log}\mathcal{V})Z_e = (\partial_I -\frac{1}{3\mathcal{V}}\tau_I)Z_e = q_I - \frac{1}{3\mathcal{V}}\tau_IZ_e \, .
\end{equation}

Let us first consider the BPS solutions. For a BPS solution from an M2 brane wrapped on a holomorphic curve we can solve $\mathcal{D}_IZ = 0$, and find
\begin{equation}\label{eqn:bpssoln}
q_I = \frac{1}{3\mathcal{V}}\tau_I Z\, .
\end{equation}
A large BPS black hole admits a solution to the equations of motion inside the K\"ahler cone, and is automatically an attractor~\cite{Chou:1997ba}. Solving the black hole equations of motion for the moduli $t$ gives the values $t_0$ fixed at the horizon. Here all BPS black holes correspond to effective curves, whose class is proportional to the self-intersection of $D$, the divisor dual to the K\"ahler form (such curves are examples of \textit{strongly movable} curves). However, not every holomorphic curve corresponds to a BPS black hole. For instance, rational curves cannot correspond to large BPS black holes, as they have negative self-intersection (or excess intersection), and the self-intersection of a K\"ahler divisor always has positive self-intersection.  This is consistent with the fact that an M2 brane wrapping a large integer multiple of a rational curve does not form a bound state and therefore does not yield a macroscopic black hole.

The entropy of a BPS black hole is written as
\begin{equation}
S^{BPS} = 2\pi \times \frac{\pi}{4 G_5} \left(\frac{1}{3} Z_e|_{t = t_0} \right)^{3/2}\, ,
\end{equation}
where the central charge is evaluated at the attractor values. It is convenient to fix $G_5 = \frac{\pi}{4}$, which we will do from now on. However, the mass of the black hole is determined by the values of the moduli at infinity, and for a BPS black hole is simply given by
\begin{equation}
M = Z_e|_{t = t_\infty}\, .
\end{equation}

Non-BPS black holes do not admit a solution to $\mathcal{D}_IZ = 0$ inside the K\"ahler cone. Instead, we must minimize the full effective potential. A critical point is given by solving
\begin{equation}
\mathcal{D}_I V_{eff} := \left(\partial_I - \frac{2}{3\mathcal{V}}\tau_I \right)V_{eff} = 0\, .
\end{equation}
However, unlike the BPS case such critical points are not guaranteed to be minima, and this must be checked example-by-example. A large non-BPS black hole then exists for a given set of charges if we can find a minimum inside the K\"ahler cone. The entropy is related to the black hole effective potential, evaluated at the attractor values:
\begin{equation}\label{eqn:nonbpsentropy}
S = 2\pi \times \left(\frac{1}{6} V_{eff}|_{t = t_0} \right)^{3/4}\, ,
\end{equation}
Again, like the BPS case the mass depends on the values of the asymptotic moduli, but generally cannot be read off as easily as in the BPS case, as when the state is not BPS, then the mass is not given by the central charge. Instead, for the most part we will fix the asymptotic values of the moduli to the attractor ones, set by the charges. Such black holes are called ``double extremal''. Here the moduli do not flow, and the mass can be read off from the minimized black hole potential~\cite{deAntonioMartin:2012bi,Meessen:2012su}, which reads
\begin{equation}
M = \sqrt{\frac{3}{2}V_{eff}}|_{t = t_0}\,.
\end{equation}

It will be equally interesting to consider black strings, obtained by wrapping an M5 brane on a divisor $D \subset X$. The procedure is analogous to the 5d black hole case~\cite{Andrianopoli:2007kz,Meessen:2012su}. The central charge for strings is given by~\cite{Chou:1997ba}
\begin{equation}
Z_m = p^I \tau_I\, ,
\end{equation}
where the $p^I$ are the quantized magnetic charges, which can be interpreted as the wrapping numbers of the M5 brane around the $I$-th divisor homology class. Similar to the black hole case, the black string effective potential takes a simple form
\begin{equation}
V_{eff}^m = 4 G_{IJ} p^I p^J\, ,
\end{equation}
where we have normalized the effective potential to be analogous to the black hole potential, so that
\begin{equation}
V_{eff}^m = \frac{2}{3}Z_m^2 + + G^{IJ}(\mathcal{D}_I Z_m)(\mathcal{D}_J Z_m)\, .
\end{equation}
Here we have
\begin{equation}
\mathcal{D}_I Z_m := (\partial_I -\frac{2}{3}\partial_I \mathrm{log}\mathcal{V})Z_m = (\partial_I -\frac{2}{3\mathcal{V}}\tau_I)Z_m = A_{IJ}p^J - \frac{2}{3\mathcal{V}}\tau_IZ_m \, .
\end{equation}
For a BPS solution we have $\mathcal{D}_I Z_m = 0$, and we find
\begin{equation}
t^I = \frac{3 p^I}{Z_m}\, .
\end{equation}
We then see that the BPS M5 branes must wrap an ample divisor. Here the solution is automatically an attractor. For a non-BPS string we must instead solve
\begin{equation}
\mathcal{D}_I V_{eff}^m := \left(\partial_I - \frac{4}{3}\tau_I\right)V_{eff}^m = 0\, .
\end{equation}
A large black string exists if these equations can be solved inside the K\"ahler cone. Similarly to the non-BPS black holes, whether the solutions are attractors must be checked case-by-case.
Like in the black hole examples we will simply set the asymptotic moduli to be the same as those fixed at the horizon $t_0$, and so we can immediately read off the black string tension as
\begin{equation}
T = \sqrt{\frac{3}{2}V_{eff}^m}|_{t = t_0}\, .
\end{equation}

\subsection{Minimal cycles and geometric measure theory}

In this section we connect the previous black hole and string formulas to geometry, in particular to computing volumes of cycles, both holomorphic and non-holomorphic. It was proposed in~\cite{Ooguri:2016pdq} that black hole physics might be able to place bounds on the volumes of non-holomorphic cycles. This idea was realized concretely in~\cite{Demirtas:2019lfi}, where the WGC was used to predict parametric recombination in a non-holomorphic class, providing the first prediction for geometric measure theory using black hole physics reasoning. In this work we develop a more direct approach, valid for ``large'' homology classes. Consider a single connected $m$-brane compactified on an $n$-cycle $\Sigma$, with $m \geq n$, leading to an $(m-n)$-brane in the non-compact space. The tension of the $(m-n)$-brane (or mass if $n = m$) is given by the volume of $\Sigma$. If we consider a single-centered black brane as we did in the previous section, we then expect the black brane tension to coincide with the volume of the cycle wrapped by the brane. This is directly verified in the BPS black brane case, where the tension is given by the central charge, computed as the volume of a holomorphic cycle of appropriate dimension. In the non-BPS black brane case we cannot compute the tension with the central charge, but the black brane effective potential computes the tension, and thus provides a prediction for the volume of the cycle. The equations of motion of the brane require that the brane wraps a cycle that corresponds to a local critical point of the volume functional on cycles in the corresponding homology class $[\Sigma]$, and to be a stable state $\Sigma$ should be a local volume-minimizer in $[\Sigma]$. If $\Sigma$ corresponds to a black brane we expect it to be a local minimizer, as (large) black brane decays are expected to follow a classically disallowed trajectory~\cite{Kraus_1995,Kraus_19952,Parikh_2000,Aalsma:2018qwy}, and admit an interpretation as a tunneling event, and we therefore expect the black brane to be a metastable configuration.

Let us then formulate the mathematical problem equivalent to our physical one. We consider a Calabi-Yau threefold $X$, with Ricci-flat metric $g$, corresponding a complex structure $\Omega$, and K\"ahler form $J$. We then consider an even-dimensional homology class $\Sigma$ of dimension $n$, which is not necessarily holomorphic. Such a cycle admits a globally voluming-minimizing representative $\Sigma_{\mathrm{min}}$, which is possibly disconnected. However, we expect the black branes to correspond instead to a connected representative, which is a local, but perhaps not global, volume minimizer. The connectedness of the representative is motivated by the fact that it is a solution to the black brane equations which represents a bound state.  We also observe that the predicted volume of the cycle is different from the volume of the disconnected, piecewise-calibrated representatives of the cycle, as we will show in numerous examples. This is similar to what one expects in the supersymmetric case.  Namely, BPS black holes correspond to strongly movable curves, which have irreducible representatives, and BPS black strings correspond to ample divisors, which also have irreducible representatives for large charge.

Denote the (possibly non-unique) connected volume minimizing representative of $\Sigma$ as $\Sigma^c_{\mathrm{min}}$ (whose identity might depend on the position in moduli space), with volume
\begin{equation}
\mathrm{vol}(\Sigma^c_{\mathrm{min}}) = \int d^n x \sqrt{|g|}_{\Sigma_{\mathrm{min}}}\, ,
\end{equation}
where $g|_{\Sigma_{\mathrm{min}}}$ is the pullback of the Ricci-flat Calabi-Yau metric to $\Sigma^c_{\mathrm{min}}$. 
The problem that we are solving by minimizing the black brane potential is to minimize $\mathrm{vol}(\Sigma^c_{\mathrm{min}})$, in the strict interior of the K\"ahler cone, subject to the constraint
\begin{equation}
\mathrm{vol}(X) = \frac{1}{3!}\int\limits_{X} J \wedge J \wedge J = 1\, .
\end{equation}

This immediately leads to a number of surprises. First, this method of computing non-holomorphic volumes does not require explicit knowledge of the Calabi-Yau metric, only the moduli. This is familiar in the BPS case when the branes wrap calibrated cycles, but is perhaps unexpected for non-calibrated cycles. Second, the complex structure moduli completely decouple, and so the computation of the non-holomorphic volume corresponding to macroscopic black brane does not depend on the complex structure moduli. This fact is then a physics proof of the following conjecture:

\begin{definition}
Consider an electric or magnetic charge, corresponding to an even-dimensional homology class $[\Sigma]$ in a Calabi-Yau threefold $X$. We assume
the class $[\Sigma]$ is ``large'' which can be accomplished for example by assuming $[\Sigma]=N[\Sigma_0]$ for an integer $N>>1$. If the corresponding black brane equations of motion are solved in the strict interior of the K\"ahler cone, and furthermore the solution is an attractor, we call the associated (locally) volume-minimizing connected representative $\Sigma$ a ``large black brane cycle'' (LBBC).
\end{definition}

\begin{conjecture}
Consider an LBBC, $\Sigma$, in a Calabi-Yau threefold $X$, and let the moduli take the corresponding attractor values $t_0$. For these values, the volume of $\Sigma$ is asymptotically independent of the complex structure moduli.  More precisely ${\rm lim}_{N\rightarrow \infty}vol([\Sigma])/N$ is independent of the complex structure moduli (and $N$).
\end{conjecture}

This should also remains true in IIa, and the same idea for 3-cycles in IIb follows from the same reasoning (and mirror symmetry). In fact, this should also imply some mirror symmetry relations between non-holomorphic cycles on one CY threefold, and non-special Lagrangian 3-cycles on the mirror.

We define an LBBC as one for which the attractor mechanism fixes the moduli strictly in the interior of the K\"ahler cone, in which case we conjecture the existence of a connected, locally volume minimizing representative of the corresponding homology class. However, for classes that do not correspond to LBBCs we do not make a statement about the existence of such a representative either way. In fact, allowing for topology changing transitions or for non-geometric phases in the flow might enlarge the space of black objects, both BPS and not, but we will not consider such effects. In the case that the minimum is located on the boundary of the K\"ahler cone we still expect the solution to be marginally under control from the string theory perspective.

We emphasize that an LBBC $\Sigma$ is a local volume minimizer of the homology class $[\Sigma]$, but is not necessarily the global volume minimizer, though in some cases we expect it to be. We will see candidates for both local and global volume minimizers in the examples. In addition, since LBBCs correspond to macroscopic black branes we expect our conclusions to hold only at asymptotically large charge; that is, for ``large'' homology classes, or large wrapping numbers. The behavior of small cycles, in the sense of homology, can and will be quite different and in  particular it could (and most likely does) depend on the complex structure moduli.

For the examples of curve LBBCs we consider, we will always be able to identify a disconnected representative of $[\Sigma]$ that has volume smaller than $\Sigma$ itself; this representative is a piecewise-calibrated representative of $[\Sigma]$, which we will denote $\Sigma^{\cup}$. Physically this simply corresponds to the fact that the non-BPS black branes can decay to BPS-anti-BPS constituents, as predicted by the  WGC. Mathematically, we are predicting the existence of a connected, locally volume minimizing representative of $[\Sigma]$, distinct from the disconnected piecewise-calibrated representatives of $[\Sigma]$.

For black strings, corresponding to M5 branes on divisors, we will observe different behavior. Via the black string tension formula for divisors we will identify examples of $\Sigma$ that have volume smaller than any $\Sigma^\cup$, making the physical prediction of a stable non-BPS string that the black string can decay into. We can always write such a class as $[\Sigma] = [\Sigma^1 - \Sigma^2]$, where both $[\Sigma^1], [\Sigma^2]$ are holomorphic, but $[\Sigma^1 - \Sigma^2]$ is not. A (non-unique) piecewise-calibrated representative of $[\Sigma]$ can always be written as $(\Sigma^1)\cup (-\Sigma^2)$. When, for all choices of piecewise-calibrated representatives, we have
\begin{equation}
\mathrm{vol}(\Sigma) < \mathrm{vol}(\Sigma^1) + \mathrm{vol}(\Sigma^2)\, ,
\end{equation}
we say the cycle $\Sigma$ has \textit{recombined}. The physical significance of recombination is that a brane wrapped on $\Sigma$ is non-BPS, but cannot completely decay into BPS-anti-BPS constituents. Recombined cycles are also associated with non-BPS instantons in $\mathcal{N} = 1$ theories, which can provide significant corrections to the K\"ahler potential~\cite{Demirtas:2019lfi}.

A demonstration of recombination is of particular general interest in geometric measure theory. There has been a great deal of progress in understanding properties of minimal cycles, such as the existence of a minimizer~\cite{FF} and the degree of singularity of the minimizer~\cite{Almgren}. However, for the Ricci-flat setting that we are most interested in, the only progress in concretely understanding recombination has been in the K3 surface. In a near-orbifold limit Micallef and Wolfson~\cite{2005math......5440M} demonstrated recombination of curves in K3 by identifying a minimal two-sphere in a class with self-intersection $-4$, which does not admit a holomorphic representative. In fact, earlier Sen~\cite{Sen:1999mg} considered the same setup from a physics perspective, and found the same result via a tachyon condensation analysis. However, Sen also found that away from the near-orbifold limit the minimal representative of the class was the piecewise-calibrated one.

The recombination results for non-holomorphic curves in K3 via Sen and Micallef-Wolfson generally seem to circumvent a direct black brane analysis. In fact, we will find no examples of curve recombination in a Calabi-Yau threefold using a black brane analysis, suggesting that if curve LBBCs do exhibit recombination, such an effect is rare. However, we do expect curve classes in Calabi-Yau threefolds that exhibit recombination more generally; a trivial example is $\mathrm{K3}\times T^2$, and less trivial is an appropriate K3 fibration to realize Sen's example. It is entirely possible though that the only curves that recombine are small curves (in the homological sense), like the Micallef-Wolfson and Sen example, and do not correspond to black holes. On the other hand, we will find divisor LBBCs for which the black string tension formula predicts recombination, suggesting the difference between the curve and divisor case is related to codimension: curves, which are codimension-two cycles in a threefold generically do not intersect, but divisors can generically intersect.  Such intersections could lead to the existence of physical modes localized there, which can condense and lead to recombination and smoothing of the BPS-anti-BPS constituents.

In any case, the black brane calculation makes a prediction for the volume of a connected representative of a non-holomorphic, even-dimensional homology class in $X$. However, this computation does not shed light on why the volume is larger or smaller than a piecewise representative of that class in any given example. In the next subsection we will propose a simple explanation for this phenomenon based on the black hole formation process. 

\subsection{Force as a recombination test}

In this section we propose a more physical test of whether recombination should or should not occur, which will lead us to a conjectural test for the possibility of recombination even when a non-holomorphic cycle is not an LBBC. 

We will use the black hole formation process, in particular the forces between the charged constituents arranged in an appropriate manner before forming the black hole, to understand whether recombination should occur. Consider decomposing the black hole into its BPS and anti-BPS constituents, separated by a large but finite distance. We will attempt to form the black hole by repeatedly bringing together a BPS constituent and an anti-BPS constituent, in the attractor background.  When the black hole has a larger mass than the sum of the masses of its BPS-anti-BPS constituents, we expect that this force is repulsive, as we will need to put energy into the system to form the black hole. We will find this is true for all but one of the black hole examples that we study, which will lead to a conjecture of recombination for small curves in that example. For black strings whose tension is greater than the sum of its parts, we will again find a repulsive force, but for black strings that exhibit recombination we will find that the force is instead attractive. 

Let us calculate the force for 5d particles. The total force between them $P$ includes gravity, force from the $U(1)$ gauge fields, and the exchange between the massless scalars, and can be written as~\cite{Lee:2018spm,Heidenreich:2019zkl}
\begin{equation}
r^3 \mathrm{vol}(S^3) P = G^{IJ}q^1_I q^2_J -G^{IJ}(\mathcal{D}_I m_1)(\mathcal{D}_J m_2) - \frac{2}{3}m_1 m_2\, , 
\end{equation}
where the $q^1_I,q^2_I$ are the charges of the two particles, and $m_1, m_2$ are the masses, which generally depend on the values of the moduli. The first term is due to the $U(1)$ gauge forces, the second to the exchange of scalars, and the third to gravity. In the case that the particles are mutually BPS then their masses are given by the central charge, and so we have $m_1 = Z_1 = q_{1}\cdot t$ and $m_2 = Z_2 = q_{2}\cdot t$, and we find that the total force vanishes between mutually BPS particles. If instead we consider a BPS particle with mass $m_1 = Z_1 = q_{1}\cdot t$ and an anti-BPS particle with mass $m_2 = -Z_2 = -q_{2}\cdot t$, we find a total force of the form
\begin{align}\label{eqn:force}
r^3 \mathrm{vol}(S^3) P 
= 2 G^{IJ}q_{1I}q_{2J}\, ,
\end{align}
which does not in general vanish. We expect that if this force is positive, energy will have to be put into the system to form an extremal black hole, and so the curve corresponding to the black hole will have greater volume than the piecewise-calibrated representative given by the BPS-anti-BPS constituents. On the other hand, if the force is negative, then the system will radiate energy as the extremal black hole forms, and so we expect black hole curve to have smaller volume than the piecewise-calibrated representative; that is, we expect non-trivial recombination.  An analogous formula holds for the force between BPS and anti-BPS strings, of the form
\begin{equation}
P \sim G_{IJ}p_1^{I}p_2^{J}\, . 
\end{equation}

Consider two effective divisors, $D_1$ and $D_2$. If we obtain a BPS string from an M5 brane on $D_1$, and an anti-BPS string from an M5 brane on $-D_2$, the force between the strings can be written as
\begin{equation}
P \sim \mathcal{V}\,\mathrm{vol}(D_1 \cap D_2) -\mathrm{vol}(D_1) \times \mathrm{vol}(D_2)\, ,
\end{equation}
where $\mathrm{vol}(D_1 \cap D_2)$ is the volume of the intersection of $D_1$ and $D_2$ when their intersection is effective, and in general is calculated as
\begin{equation}
\mathrm{vol}(D_1 \cap D_2) := D \cdot D_1 \cdot D_2\, ,
\end{equation}
where $D$ is the K\"ahler divisor. In particular, we see that for non-intersecting $D_1$ and $D_2$, the force between the BPS and anti-BPS constituents is always attractive:
\begin{equation}
P \sim -\mathrm{vol}(D_1) \times \mathrm{vol}(D_2)\, .
\end{equation}

The Micallef-Wolsfon and Sen analyses of recombination in K3 provide instructive examples. In both setups, the curve under consideration was a $(-4)$-curve in a non-holomorphic class $[\Sigma] = [\Sigma_1 - \Sigma_2]$, where both $\Sigma_1$ and $\Sigma_2$ are holomorphic curves, and $\Sigma_1 \cap \Sigma_2 =0$. If we also
compactify on a $T^2$ to arrive at a 5d theory, the force between a BPS M2 brane on $\Sigma_1$ and an anti-BPS M2 brane on $-\Sigma_2$ is
\begin{equation}
P \sim -\mathrm{vol}(\Sigma_1)\times \mathrm{vol}(\Sigma_2)\, ,
\end{equation}
which is attractive everywhere inside the K\"ahler cone. This is indicative that $\Sigma_1$ and $-\Sigma_2$ might recombine. However, whether or not they actually recombine depends on other moduli. While $\Sigma_1 \cap \Sigma_2 =0$, there is another class $\Sigma_3$ such that $\Sigma_1 \cdot \Sigma_3 = \Sigma_2 \cdot \Sigma_3 = 1$. Let the volume of $\Sigma_3$ be $\epsilon$. Both Micallef-Wolfson and Sen showed that in the case that $\epsilon \ll 1$, the class $\Sigma$ exhibited recombination; however, Sen showed that when $\epsilon$ become larger, and the K3 moved away from a near-orbifold limit, the volume-minimizer of the class $[\Sigma]$ was the disconnected, piecewise-calibrated representative $(\Sigma_1) \cup (-\Sigma_2)$. In that case, the attractive force does not indicate recombination, but instead the presence of a bound state of particles in the non-compact spacetime. Therefore, we expect that the force test to be a check for the possibility of recombination, but not a guarantee, as it might indicate a bound state instead.

\section{Black Hole Examples}\label{sec:bhexamples}
We will now explore some examples of 5d black holes, via M2 branes wrapped on curves in Calabi-Yau threefolds, discussing aspects of measure theory and entropy throughout. We will find that none of the examples studied by black hole techniques exhibit recombination, but instead the LBBC corresponds to a connected local, but not global, volume-minimizing representative of the corresponding homology class. The cases we examine are some of those for which we know the integral generators of the semigroup of effective (holomorphic) curves.

In all of the Calabi-Yau threefolds that we consider we will identify the curves that correspond to large black holes, both BPS and non-BPS. We will plot these regions, where yellow denotes the region of macroscopic BPS black holes, blue denotes the region of macroscopic non-BPS black holes, and white denotes the region where no macroscopic black holes exist.   Note that the first quadrant in these plots always corresponds to the cone generated by holomorphic curves, as the positive $x$ and $y$ axes correspond to the generators of the semigroup of holomorphic curves.  An example of such a plot is given in Fig.~\ref{fig:P3P1bh}, which is the first example that we explore.  We will see that the BPS black holes always correspond to a proper subset of holomorphic curves.  These regions without macroscopic black holes but with holomorphic curves are familiar from examples such as the conifold.   We do not find any overlap of the BPS and non-BPS black hole regions in our examples.

\subsection{Hypersurfaces in smooth Fano toric fourfolds}

We will start with examples realized as hypersurfaces in smooth Fano toric varieties. In these examples we can infer the generators of the semigroup of effective curves in  the Calabi-Yau $X$ from those of the ambient toric variety $V$. In~\cite{2019arXiv191103146S} the integral Hodge conjecture was proved for anticanonical hypersurfaces of smooth Fano toric $n$-folds, with $n\geq 4$. In particular for our purposes, it was shown in this work that the integral generators of the semigroup of algebraic curves in $Y$ all have algebraic representatives in $X$, and that these curves generate the semigroup of algebraic curves in $X$. Therefore, to find the integral effective curves in $X$ we need only to perform the task for the ambient space $Y$, for which there is a well-known algorithm, as the semigroup is generated by the torus-invariant curves~\cite{cox2011toric}.

We will now focus on minimizing the effective potential for an M2 brane wrapping an arbitrary, non-calibrated curve in $X$, where $X$ is a Calabi-Yau threefold. Let us start with a simple example we can solve exactly, namely an $h^{1,1} = 2$, K3 fibered example.

\subsubsection{A K3 Fibration}\label{sec:k3bh}

We consider a generic anticanonical hypersurface $X \subset \mathbb{P}^3 \times \mathbb{P}^1$, with corresponding hyperplane divisor classes $D_1$ and $D_2$, given by the ambient hyperplane classes restricted to the hypersurface. This example is $K3$ fibered, with the typical fiber given by the class $D_2$.  Expanding the K\"ahler form as\footnote{Indices on $t^I$ are raised and lowered with a Kronecker delta function, which we use for ease of presentation.}
\begin{equation}
J = t_1 [D_1] + t_2 [D_2]\, ,
\end{equation}
where $[\dots]$ denotes the dual $(1,1)$-form, the volume of $X$ takes the form
\begin{equation}
\mathcal{V} = \frac{1}{3}t_1^2 ( t_1+ 6 t_2)\, .
\end{equation}
As the ambient space is smooth Fano, the K\"ahler cone of the hypersurface is inherited from the ambient space, and the K\"ahler cone conditions are that the parameters $t_I$ are positive.

Let us first determine the charges $q_I$. Denoting divisors in the ambient space $V = \mathbb{P}^3 \times \mathbb{P}^1$ with hats, the cone of curves in $V$ is generated by
\begin{equation}
 C_1 = \hat{D}_2^3\, , \quad   C_2 = \hat{D}_2^2 \hat{D}_1\,.
\end{equation}
$C_2$ is actually proportional to the complete intersection in $X$ of $D_1$ and $D_2$. One way to realize $C_1$ in $X$ is to specialize the hypersurface equation $F = 0$ to have no $x_4^4$ term, where the $x_\alpha$, $\alpha = 1, 2,3,4$ are the projective coordinates on $\mathbb{P}^3$. In this case setting $x_1 = x_2 = x_3 = 0$ gives a $\mathbb{P}^1$ in $V$ that automatically lies in $X$, and is the base of the K3 fibration. Regardless of any specialization, $C_1$ is the base $\mathbb{P}^1$ of the K3 fibration, and $C_2$ the fibral curve.

For a general curve class $C = \alpha C_1 + \beta C_2$, the charges can be computed as
\begin{align}
&q_1 = D_1 \cdot C = \hat{D}_1 \cdot (\alpha C_1 + \beta C_2) = \beta\, ,\nonumber \\
&q_2 = D_2 \cdot C = \hat{D}_2 \cdot (\alpha C_1 + \beta C_2) = \alpha\,.
\end{align}
We have
\begin{equation}
V_{eff} = -\frac{1}{3} \alpha  \beta  t_1^2+\beta ^2 t_1^2+\frac{1}{12} \alpha ^2 \left(t_1^2+8t_1 t_2+24 t_2^2\right)\, ,
\end{equation}
subject to the constraint $\mathcal{V} = 1$.  Define
\begin{equation}
x = \frac{t_1}{t_2}\, .
\end{equation}
The conditions that $\mathcal{D}_I V_{eff} = 0$ can then be written as
\begin{equation}\label{eqn:k3eom}
(\alpha  (x+4)-2 \beta  x) (\alpha  (x+12)+6 \beta  x)=0\, .
\end{equation}

Let us first determine the conditions for a BPS black hole, which corresponds to the vanishing of the first factor in Eq.~\ref{eqn:k3eom}. We have
\begin{equation}
x = \frac{4 \alpha }{2\beta - \alpha}\, ,
\end{equation}
and enforcing that $\mathcal{V} = 1$ we find
\begin{equation}
\frac{1}{3}x^2 (6 +x) t_2^3 = 1\, .
\end{equation}
In order to have $t_1,t_2 > 0$ we need to enforce that $x>0, t_2>0$, which leads to the conditions
\begin{equation}
\{\beta < 0\,\,\, \mathrm{ and } \, \, \, 2\beta < \alpha < 0\} \quad \mathrm{or} \quad \{\beta > 0\,\,\, \mathrm{ and } \, \, \, 2 \beta > \alpha > 0\}\, .
\end{equation}
The entropy of the BPS black holes is then written as
\begin{equation}
S = \frac{\pi \sqrt{\alpha}(6\beta - \alpha)}{6\sqrt{2}}\, .
\end{equation}

Let us next determine the conditions for a non-supersymmetric black hole, which corresponds to the vanishing of the second factor in Eq.~\ref{eqn:k3eom}. We have
\begin{equation}
x =  -\frac{12 \alpha }{\alpha +6 \beta }\, ,
\end{equation}
Again, in order to have $t_1,t_2 > 0$ we need to enforce that $x>0, t_2>0$, which leads to the conditions
\begin{equation}\label{eqn:nonBPScharges}
\{\beta < 0\,\,\, \mathrm{ and } \, \, \, 0 < \alpha < -6\beta\} \quad \mathrm{or} \quad \{\beta > 0\,\,\, \mathrm{ and } \, \, \, -6 \beta < \alpha < 0\}\, .
\end{equation}
These solutions all correspond to attractors. The entropy of the non-BPS black hole takes the form
\begin{equation}
S = \frac{\pi \sqrt{|\alpha|}|6\beta - \alpha|}{6\sqrt{2}}\, ,
\end{equation}
which is simply the BPS entropy with appropriate sign flips.

To examine the mass of the non-BPS black holes, define the ratio of the non-BPS black hole mass to the minimal piecewise-calibrated representative of the corresponding homology class as $R$. In this example we have, in the non-BPS black hole region of parameters,
\begin{equation}
R = 1 - \frac{4\alpha}{\alpha + 18\beta}\, ,
\end{equation}
which is always larger than unity in this region. Therefore these black holes are unstable to decay to their holomorphic-anti-holomorphic constituents. We conclude that the non-BPS black holes correspond to M2 branes wrapping connected representatives of the corresponding homology class which are local, but not global, volume minimizers.

\begin{figure}
\begin{center}
  \includegraphics{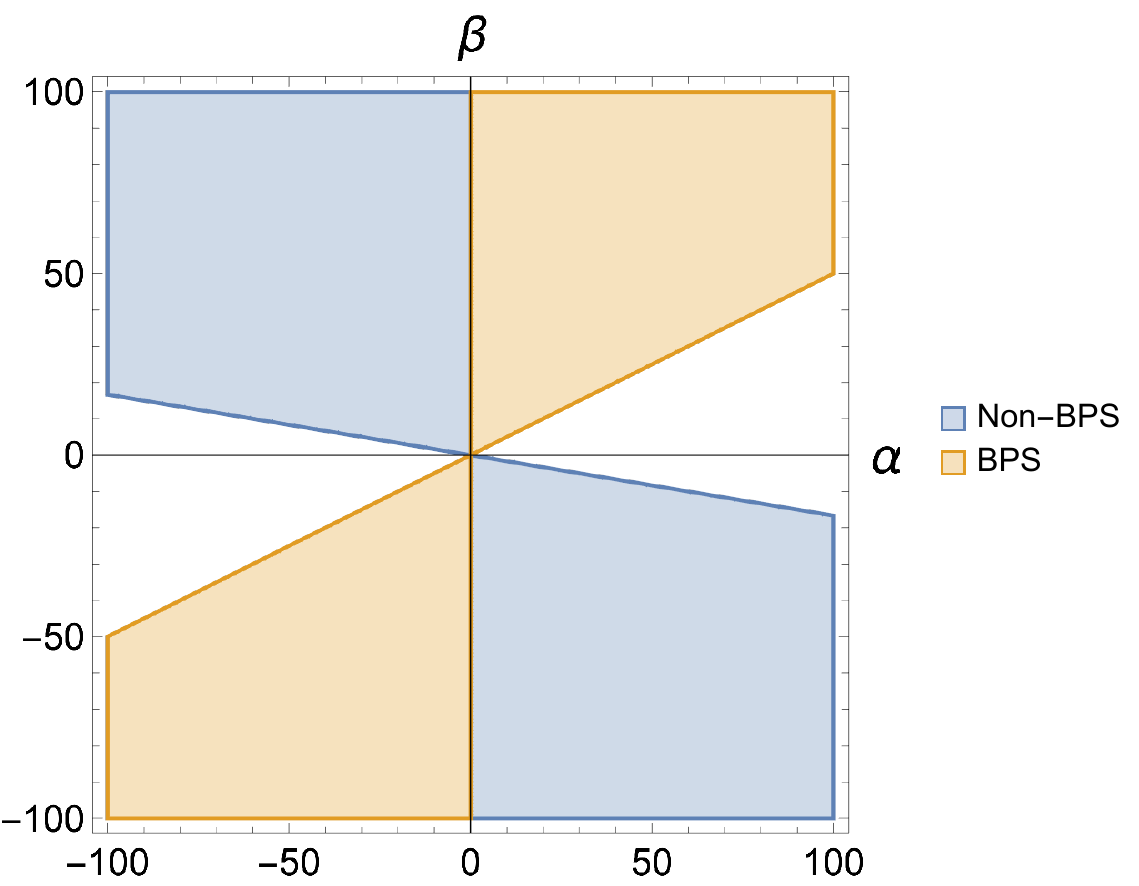}
\caption{The regions of curve space with large black holes for a generic CY hypersurface in $\mathbb{P}^3 \times \mathbb{P}^1$. The axes indicate the homology class, specified by $\alpha$  and $\beta$. The white region indicates that there is no large black hole for that particular homology class.}\label{fig:P3P1bh}
\end{center}
\end{figure}

In Fig.~\ref{fig:P3P1bh} we show the regions where large black holes exist for both the BPS and non-BPS cases. Note that certain curve homology classes, namely the white regions, do not correspond to large black holes.

While it is not the main focus of this work, for this particular example we can determine a \textit{fake superpotential}~\cite{Freedman:2003ax,Celi:2004st,Zagermann:2004ac,Skenderis:2006jq,Ceresole:2007wx,Gendler:2020dfp},  denoted $\mathcal{W} \neq Z$, that satisfies
\begin{equation}
V_{eff} = \frac{2}{3}\mathcal{W}^2+ G^{IJ}\mathcal{D}_I \mathcal{W}\mathcal{D}_J \mathcal{W}\, ,
\end{equation} 
whose critical points $\mathcal{D}_IW = 0$ then give non-BPS black holes. If the correct fake superpotential can be identified, the mass of of the black hole can be read off for asymptotic moduli that differ from the attractor values:
\begin{equation}
M = \mathcal{W}(t_\infty) \, .
\end{equation}
To find $\mathcal{W}$, we can make a linear transformation on the K\"ahler parameters to bring the volume to a factorized form, where the gauge kinetic function is diagonal, which will allow us to read off the fake superpotential~\cite{Gendler:2020dfp}. Defining
\begin{equation}
t_3 = \frac{1}{3}(t_1 + 6t_2)\, ,
\end{equation}
we have
\begin{equation}
t_2 = \frac{1}{6}(3 t_3 - t_1)\, .
\end{equation}
The volume then takes the form
\begin{equation}
\mathcal{V} = t_1^2 t_3\, .
\end{equation}
To read off the charges in the new basis, we note that the central charge is invariant under the coordinate transformation. We then require
\begin{equation}
Z = t_1 q_1 + t_2 q_2 = t_1 q_a + t_3 q_b\, .
\end{equation}
This leads to
\begin{equation}
q_a = q_1 - \frac{q_2}{6} = \beta - \frac{\alpha}{6} \, , \quad q_b = \frac{q_2}{2}= \frac{\alpha}{2}\, .
\end{equation}
In the basis K\"ahler moduli basis of $\{t_1,t_3\}$ the gauge kinetic function is diagonal, and so we can immediately read off the fake superpotential as
\begin{equation}
\mathcal{W} = t_1 |q_a| + t_3 |q_b|\, .
\end{equation}
Transforming back to the original geometric basis, we have
\begin{equation}\label{eqn:P3P1w}
\mathcal{W} = t_1 |q_1 - \frac{q_2}{6}| + \frac{1}{6}(t_1 + 6 t_2)|q_2| = t_1 \left| \beta -\frac{\alpha }{6}\right| +\frac{1}{6} | \alpha |  (t_1+6
   t_2)\, .
\end{equation}
It is straightforward to verify that
\begin{equation}
V_{eff} = \frac{2}{3}\mathcal{W}^2+ G^{IJ}(\mathcal{D}_I \mathcal{W})(\mathcal{D}_J \mathcal{W})\, .
\end{equation}

Let us now examine the volume predicted by this fake superpotential. We can compare the predicted volume in Eq.~\ref{eqn:P3P1w} to that of a piecewise-calibrated representative of this class. The unique volume minimizing piecewise-calibrated representative has a volume given by
\begin{equation}
\mathrm{vol}^{\cup}(\alpha,\beta) = t_1|\beta| + t_2 |\alpha|\, ,
\end{equation}
Inside the K\"ahler cone we always have $\mathrm{vol}_V^{\cup}(\alpha,\beta) < \mathcal{W}$ for the region of non-BPS charges given in Eq.~\ref{eqn:nonBPScharges}, and so for any choice of the asymptotic moduli (subject to consistency of the solution along the flow), this black hole is unstable to decay to its holomorphic-anti-holomorphic constituents.

Let us perform the force test. The total force $P$ (including gravitational, scalar, and gauge) is given in Eq.~\ref{eqn:force}, which in this example can be written as
\begin{equation}
P \sim -\frac{1}{3}t_1^2 \alpha \beta\, ,
\end{equation}
which is repulsive in the non-BPS case, where $\alpha$ and $\beta$ are of opposite sign. Therefore, to form a black hole one need to put additional energy into the system to bring the particles together, and so the mass of the black hole is larger than the masses of the constituent particles. This is consistent with the representative corresponding to the black hole being a local, but not global, volume minimizer.

\subsubsection{The bi-cubic in $\mathbb{P}^2 \times \mathbb{P}^2$}\label{sec:p2p2bh}

We now consider a generic anticanonical hypersurface $X \subset \mathbb{P}^2 \times \mathbb{P}^2 := V$. Let $\hat{D}_1,\hat{D}_2$ correspond to the hyperplane sections of each $\mathbb{P}^{2}$ factor, and let $D_1,D_2$ denote the hyperplanes restricted to $X$. Expanding the K\"ahler form as $J = t_1 [D_1] + t_2 [D_2]$, the volume of $X$ takes the form
\begin{equation}
\mathcal{V} = \frac{3}{2} t_1 t_2 (t_1+t_2)\, .
\end{equation}
 The semigroup of effective curves in $V$, and thus $X$, is generated by $C_1 :=\hat{D}_1^2 \cdot \hat{D}_2$ and $C_2 := \hat{D}_2^2 \cdot \hat{D}_1$. A general curve $C$ can then be expressed as
\begin{equation}
C = \alpha C_1 + \beta C_2\, ,
\end{equation}
and the charges are
\begin{align}
& q_1 = D_1 \cdot C = \hat{D}_1 \cdot ( \alpha \hat{D}_1^2\cdot \hat{D}_2 + \beta \hat{D}_2^2\cdot \hat{D}_1) = \beta\, , \nonumber \\
 &q_2 = D_2 \cdot C = \hat{D}_2 \cdot ( \alpha \hat{D}_1^2\cdot \hat{D}_2 + \beta \hat{D}_2^2\cdot \hat{D}_1) = \alpha\,. 
\end{align}
We have
\begin{equation}
V_{eff} = \frac{\alpha ^2 t_2^2 \left(2 t_1^2+2 t_1
   t_2+t_2^2\right)-2 \alpha  \beta  t_1^2 t_2^2+\beta ^2
   t_1^2 \left(t_1^2+2 t_1 t_2+2
   t_2^2\right)}{t_1^2+t_1 t_2+t_2^2}\, ,
\end{equation}
subject to the constraint $\mathcal{V} = 1$.  Again define
\begin{equation}
x = \frac{t_1}{t_2}\, .
\end{equation}
Let us first determine the conditions for a supersymmetric black hole, which corresponds to 
\begin{equation}
\alpha +2 \alpha  x=\beta  x (x+2)\, .
\end{equation}
Solving this gives 
\begin{equation}
    x= \frac{\sqrt{\alpha ^2-\alpha  \beta +\beta ^2}+\alpha -\beta }{\beta
   }\, ,
\end{equation}
and enforcing that $\mathcal{V} = 1$ we find
\begin{equation}
\frac{3}{2}x (1 +x) t_2^3 = 1\, .
\end{equation}
In order to have $t_1,t_2 > 0$ we need to enforce that $x>0, t_2>0$, which leads to the conditions
\begin{equation}
\{\beta < 0\,\,\, \mathrm{ and } \, \, \, \alpha < 0\} \quad \mathrm{or} \quad \{\beta > 0\,\,\, \mathrm{ and } \, \, \, \alpha > 0\}\, .
\end{equation}

We now calculate the entropy. Define $y = \alpha/\beta$. The BPS black hole entropy takes the form
\begin{equation}
S = \frac{4}{9} \pi  \left(\frac{\beta ^2 \left(y \left(5 y+4 \sqrt{(y-1) y+1}-5\right)-2
   \sqrt{(y-1) y+1}+2\right)}{\left(4 y \left(y+\sqrt{(y-1) y+1}-1\right)-2 \sqrt{(y-1)
   y+1}+2\right)^{2/3}}\right)^{3/4}\, .
\end{equation}
In the limit that $\alpha \gg \beta$, the entropy becomes
\begin{equation}
S \simeq \sqrt{\frac{2}{3}} \pi  \beta\sqrt{\alpha } \, ,
\end{equation}
while in the limit $\beta \gg \alpha$, the entropy becomes
\begin{equation}
S \simeq \sqrt{\frac{2}{3}} \pi  \alpha  \sqrt{\beta }\, .
\end{equation}

\begin{figure}
\begin{center}
  \includegraphics{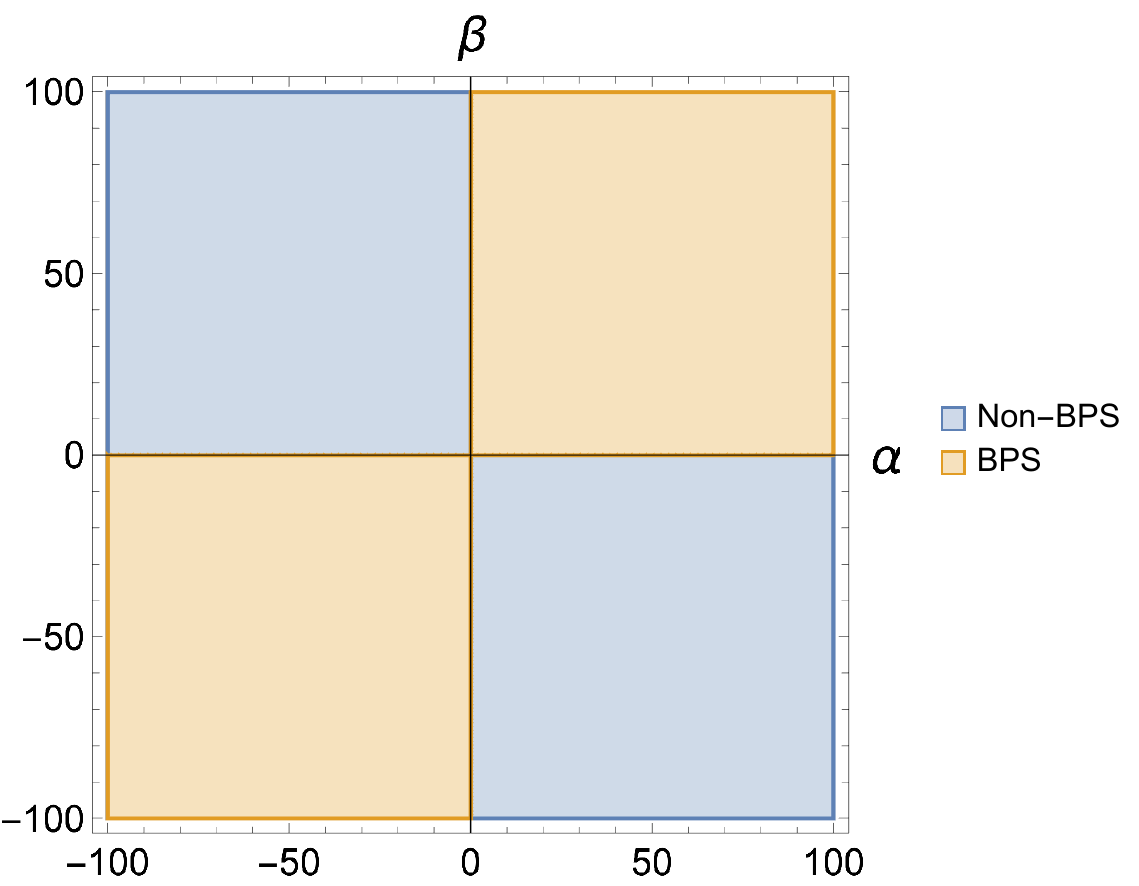}
\caption{The regions of curve space with large black holes, for a generic CY hypersurface in $\mathbb{P}^2 \times \mathbb{P}^2$. The axes indicate the homology class, specified by $\alpha$  and $\beta$. All curves, except $\alpha = 0$ or $\beta = 0$, correspond to large black holes.}\label{fig:P2P2bh}
\end{center}
\end{figure}

Let us next determine the conditions for a non-supersymmetric black hole, given by a solution to
\begin{equation}\label{eqn:quinticpoly1}
2\beta  x^5+ x^4 (4 \alpha +3 \beta )+x^3 (8 \alpha +7 \beta )+x^2 (7 \alpha +8 \beta )+x (3 \alpha +4 \beta )  + 2 \alpha =0\, .
\end{equation}
Again, in order to have $t_1,t_2 > 0$ we need to enforce that $x>0, t_2>0$. Eq.~\ref{eqn:quinticpoly1} has a positive root if
\begin{equation}
\{\beta < 0\,\,\, \mathrm{ and } \, \, \, \alpha > 0\} \quad \mathrm{or} \quad \{\beta > 0\,\,\, \mathrm{ and } \, \, \, \alpha < 0\}\, ,
\end{equation}
which covers all non-holomorphic curves. This is evident from the fact that the constant term is $2 \alpha$, and the highest power monomial coefficient is $2 \beta$, and so if $\alpha$ and $\beta$ are of differing sign there must be a zero for positive $x$. These solutions all correspond to attractors.
In Fig.~\ref{fig:P2P2bh} we show the regions where large black holes exist for both the BPS and non-BPS cases. In this example all curve classes correspond to large black holes, except when $\alpha =0$ or $\beta = 0$.

\begin{figure}
\begin{center}
  \includegraphics{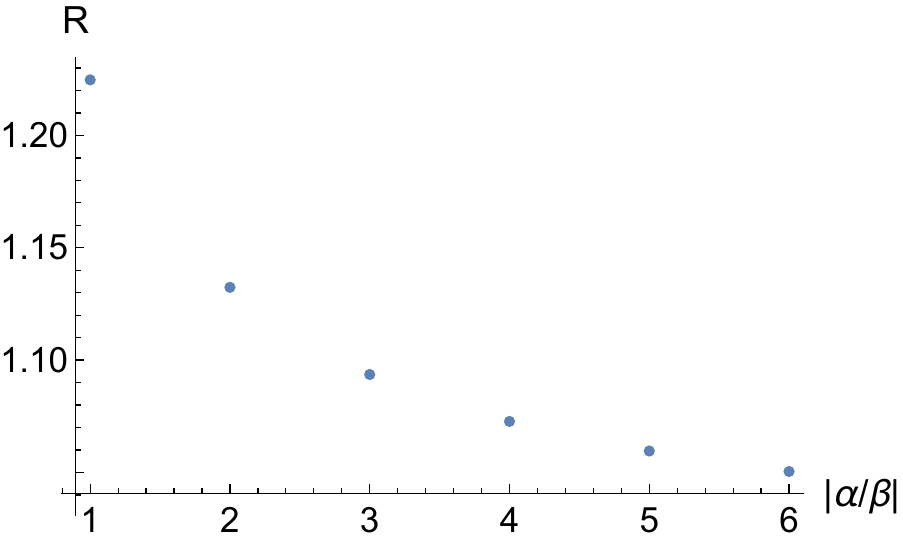}
\caption{The ratio of the black hole mass to the volume of the minimal piecewise-calibrated representative, as a function of $\alpha/\beta$, for a generic CY hypersurface in $\mathbb{P}^2 \times \mathbb{P}^2$. This ratio is always greater than unity, but approaches unity as the homology class becomes ``mostly'' holomorphic, and the corresponding black hole solution approaches a BPS one.}\label{fig:P2P2bhratio}
\end{center}
\end{figure}

Let us examine the mass of the non-BPS black holes. We cannot solve Eq.~\ref{eqn:quinticpoly1} analytically, so we will instead investigate it numerically. Taking $r = \alpha/\beta$, we plot the ratio $R$ of the black hole mass to the piecewise-calibrated volume of the corresponding cycle in Fig.~\ref{fig:P2P2bhratio}, with the asymptotic moduli set to the attractor values. We again find $R >1$ in the non-BPS case.  Therefore these black holes are unstable to decay to their BPS-anti-BPS constituents, and we conclude that the non-BPS black holes correspond to M2 branes wrapping connected representatives of the corresponding homology class which are local, but not global, volume minimizers.

The entropy can be calculated numerically as well for any given set of charges. Let us compute the entropy in the case that one of the charges is much larger in magnitude than the other, e.g., $|\alpha| \gg |\beta|$. In this case, we have
\begin{equation}\label{eqn:quinticpoly3}
2\beta  x^5+ 4 \alpha x^4 +8 \alpha x^3 +7 \alpha  x^2 +3 \alpha x  + 2 \alpha =0\, .
\end{equation}
Taking $x \sim \alpha$, we consistently find
\begin{equation}
x = -\frac{2\alpha}{\beta}\, ,
\end{equation}
for which we find an entropy of the form
\begin{equation}\label{eq:p2bigentropy}
S \simeq \sqrt{\frac{2}{3}} \pi  |\beta|\sqrt{|\alpha|}\, .
\end{equation}
This entropy takes the same form as the BPS entropy, with appropriate sign flips in the charges. Via symmetry the same hold for $|\beta| \gg |\alpha|$, with $\alpha \leftrightarrow \beta$ in Eq.~\ref{eq:p2bigentropy}.

However, this matching to the BPS entropy is only true when one of the charges is much larger in magnitude than the other. For instance, taking $\alpha = \beta$, in the BPS case, we find an entropy of
\begin{equation}
S = 1.97 \beta^{3/2}\, ,
\end{equation}
while if we take $\alpha = -\beta$, in the non-BPS case, we find an entropy of
\begin{equation}
S = 2.68 \beta^{3/2}\, .
\end{equation}

We perform the same physical test of recombination as in the previous example, where we set the moduli to the attractor values, and decompose the black hole into its BPS and anti-BPS constituents, separated by some finite distance. Again, the force, $P$, between the constituents is repulsive inside the K\"ahler cone for mixed sign $\alpha$ and $\beta$:
\begin{equation}
P \sim -\frac{2t_1^2 t_2^2 \alpha \beta}{t_1^2 + t_1 t_2 + t_2^2}\, , 
\end{equation}
and so to form the black hole energy must be added to the system, again indicating the connected representative of the corresponding homology class of the black hole is locally, but not globally, voluming minimizing.

\subsubsection{Other hypersurfaces}
We will briefly discuss general features observed in performing the same analysis for other Calabi-Yau hypersurfaces in smooth Fano toric fourfolds. Let us start with the simplest non-trivial case, which is the case of Picard rank two. 

\begin{figure}
\begin{tabular}{ccc}
  \includegraphics[width=50mm]{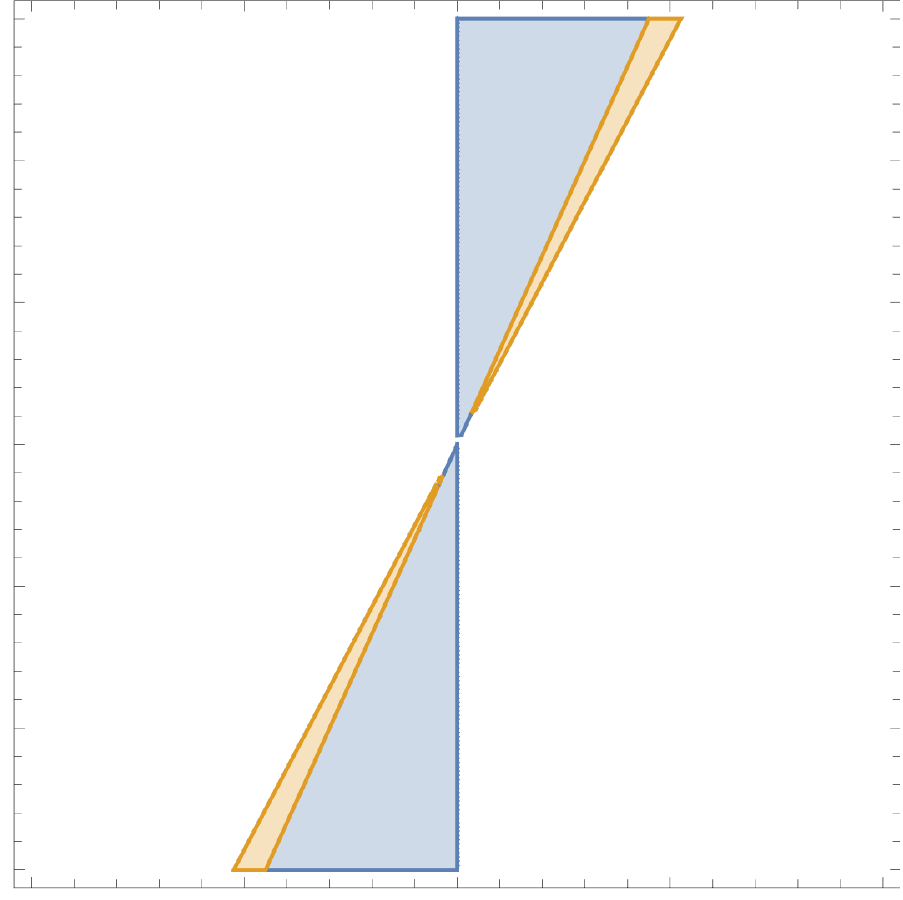} &   \includegraphics[width=50mm]{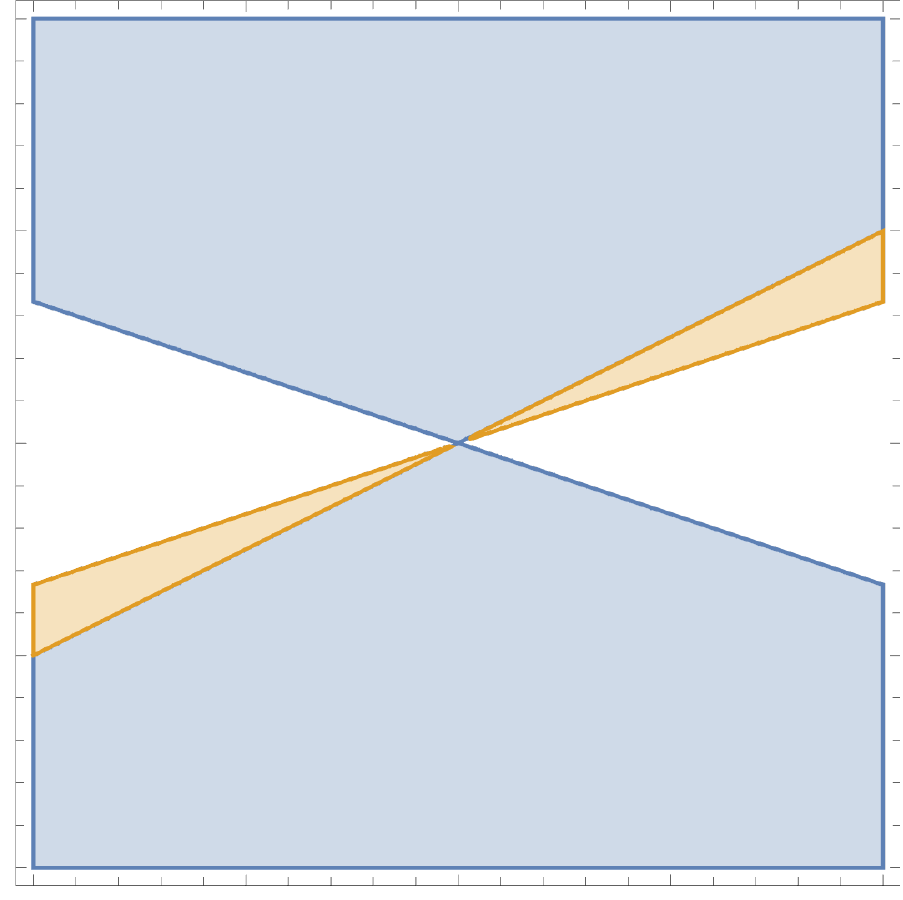} &   \includegraphics[width=50mm]{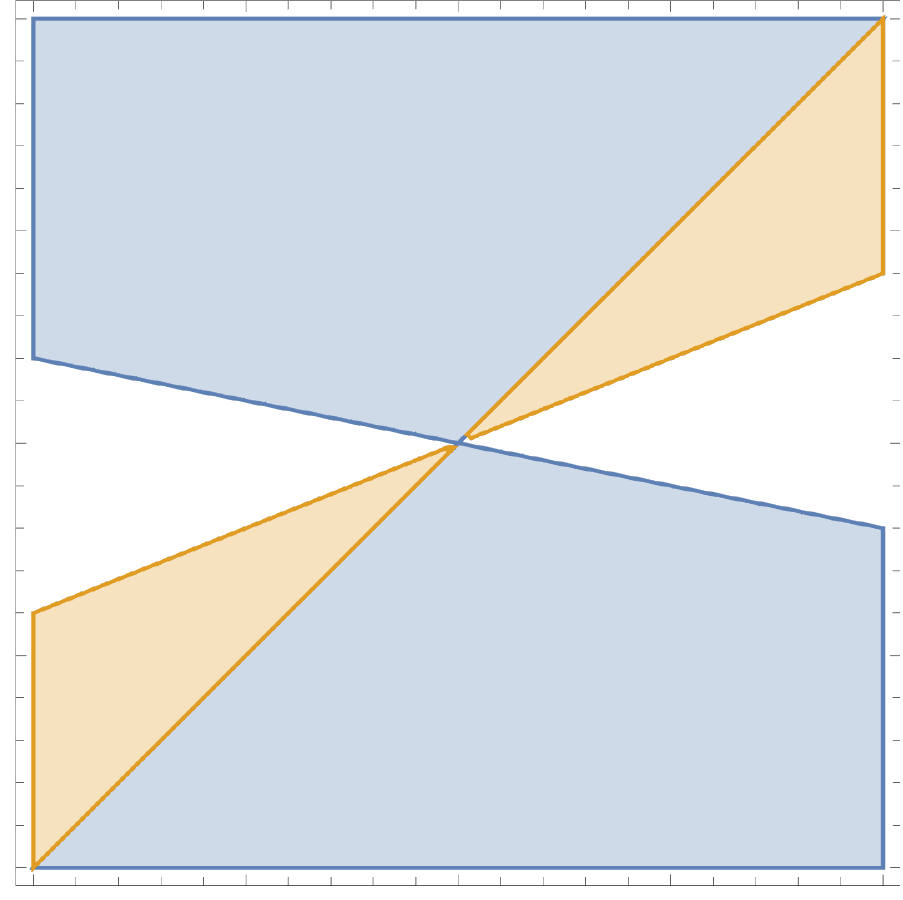} \\
(a) first & (b) second & (c) third\\[6pt]
 \includegraphics[width=50mm]{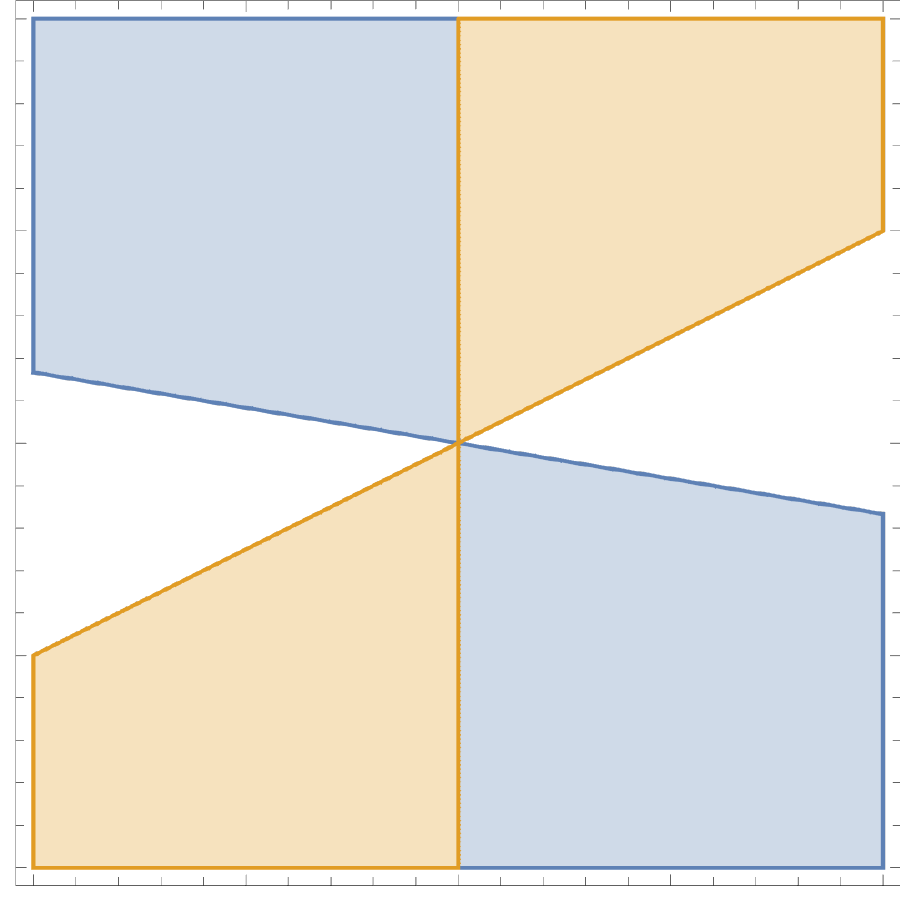} &   \includegraphics[width=50mm]{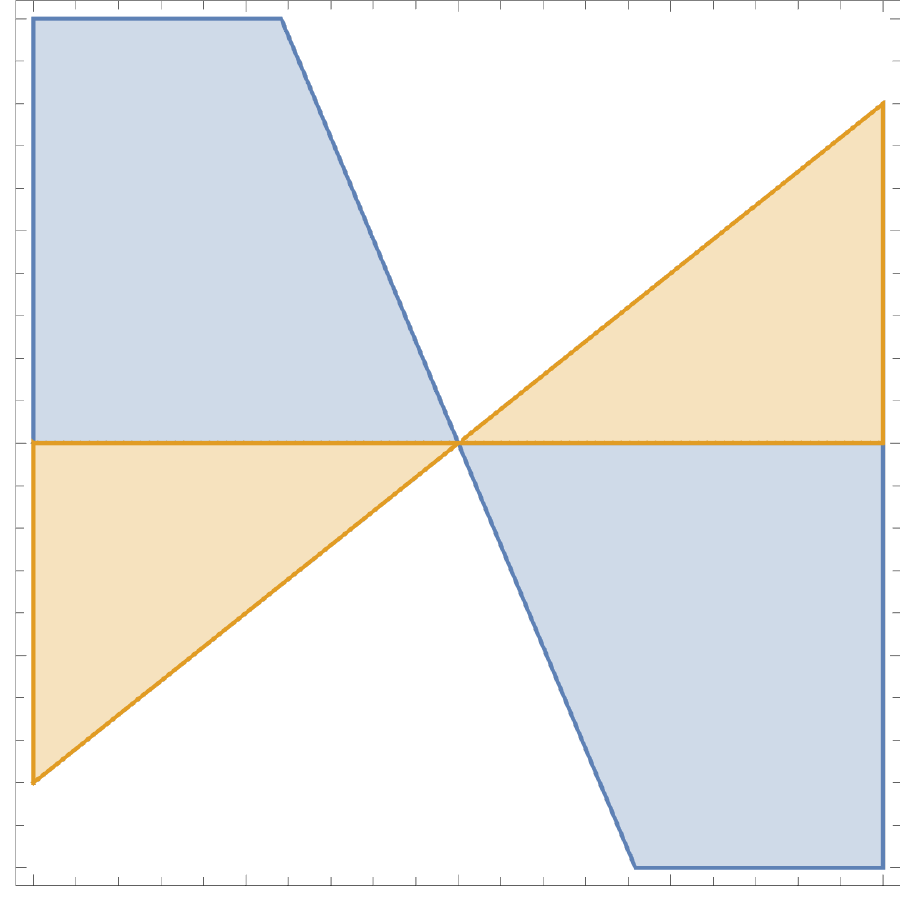} &   \includegraphics[width=50mm]{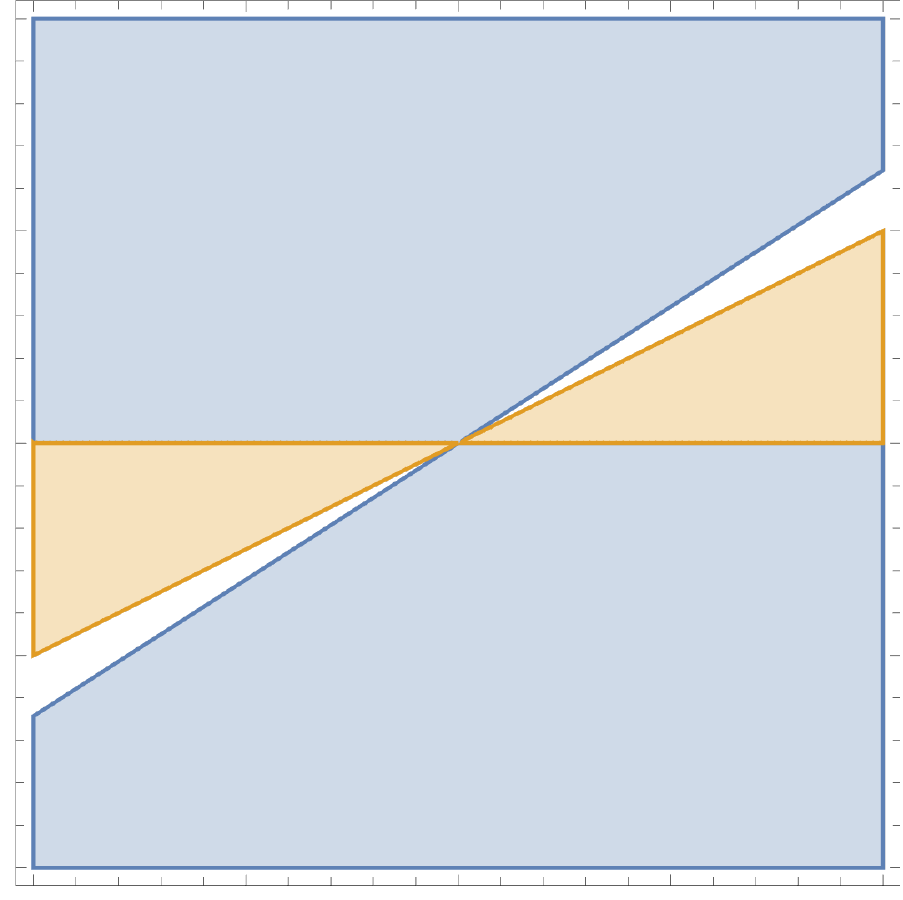} \\
(c) fourth & (d) fifth & (e) sixth\\[6pt]
\includegraphics[width=50mm]{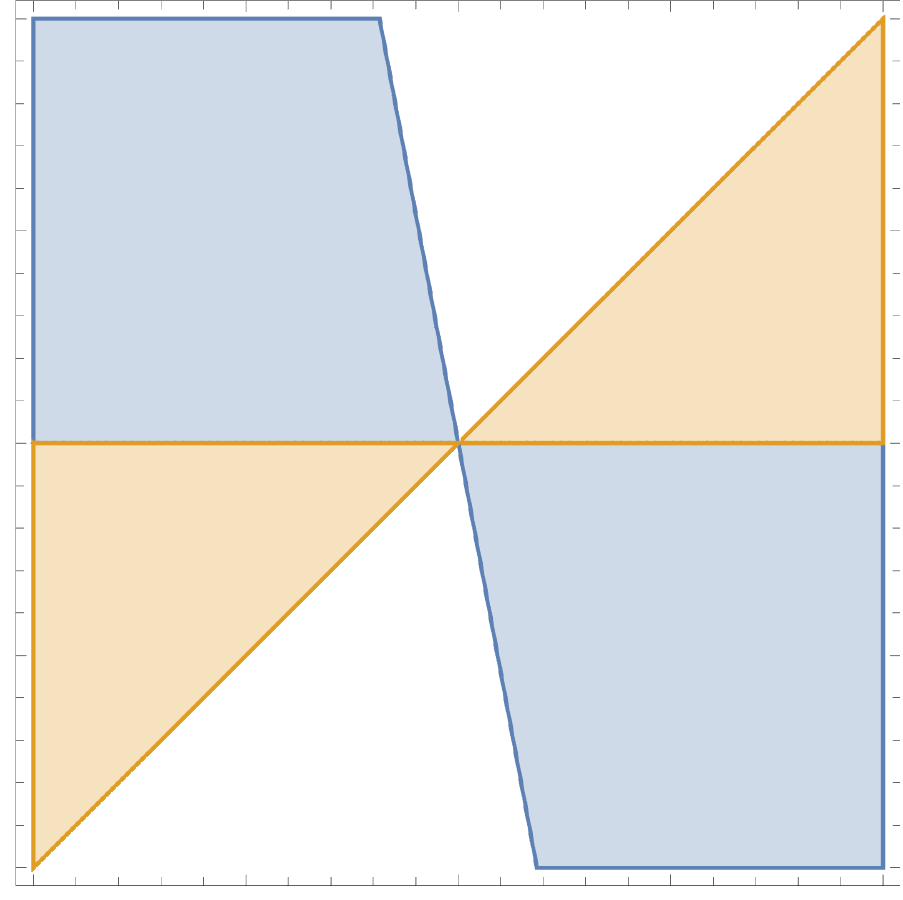} &   \includegraphics[width=50mm]{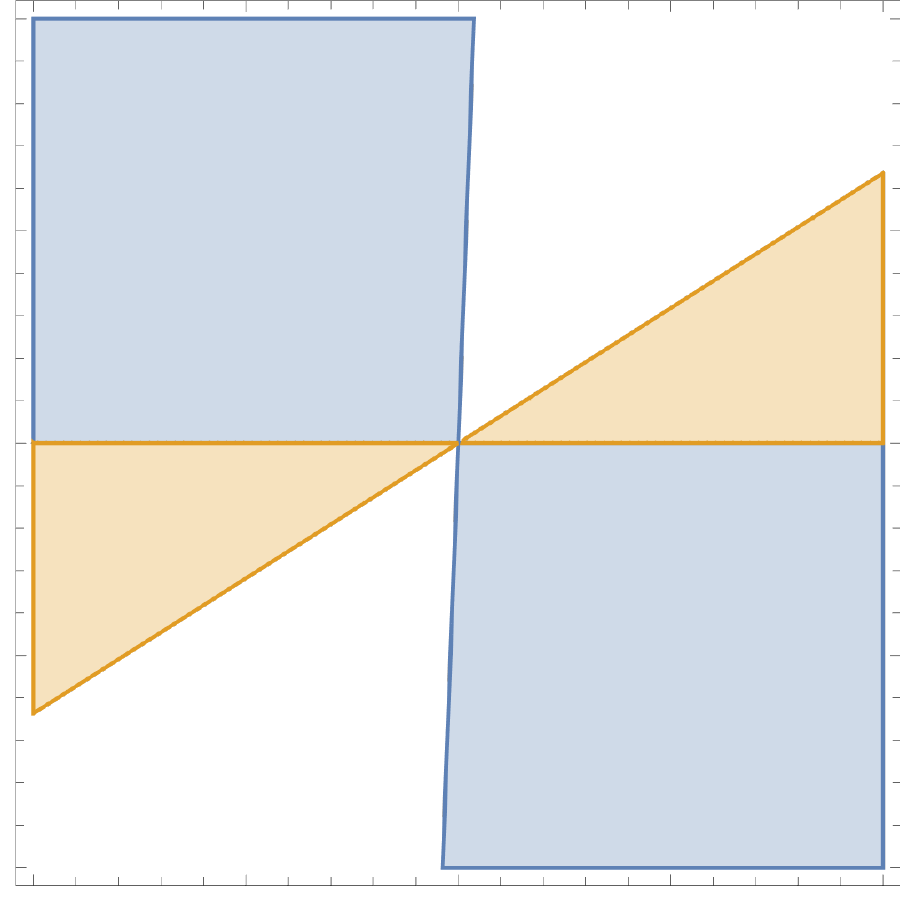} &   \includegraphics[width=50mm]{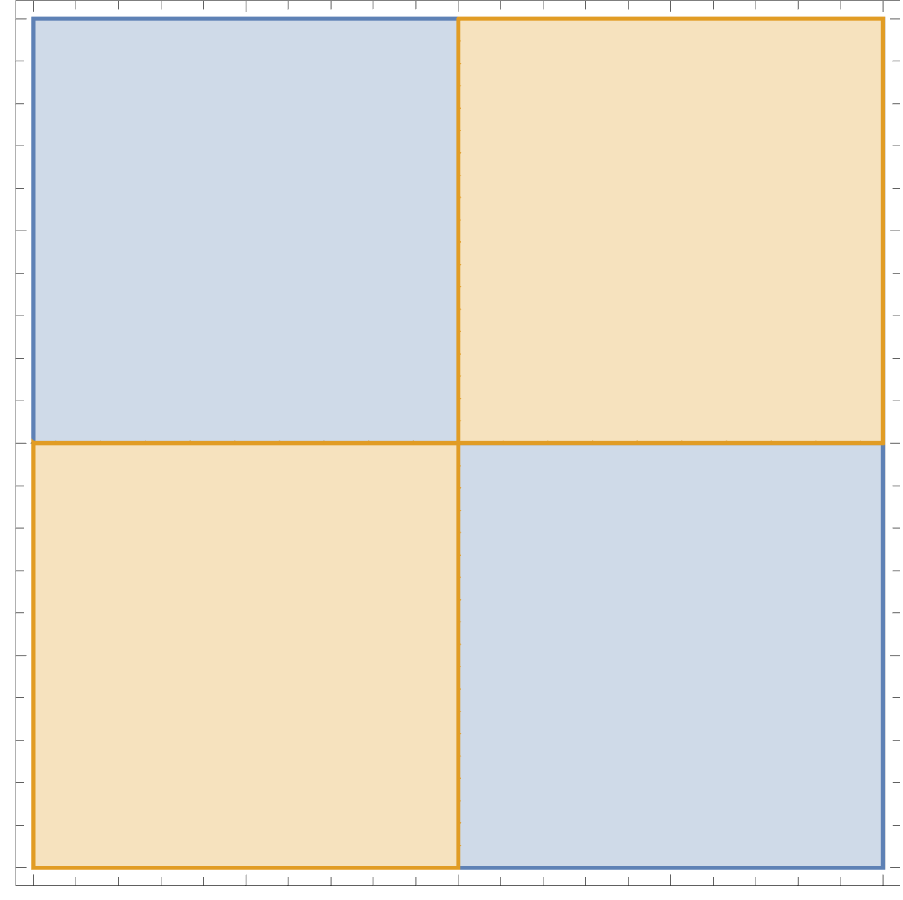} \\
(c) seventh & (d) eigth & (e) ninth\\[6pt]
\end{tabular}
\caption{The regions of curve space with large black holes, in the nine $h^{1,1} = 2$ models corresponding to anticanonical hypersurfaces in smooth Fano toric fourfolds. In each case the Mori cone (cone of effective curves) corresponds to the positive quadrant. The yellow indicates BPS black holes, the blue non-BPS black holes, and the white region indicates where a large black hole does not exist. The fourth and ninth examples correspond to the examples studied above.}\label{fig:allpicardtwo}
\end{figure}

Including the two previous examples, there are nine smooth Fano toric fourfolds with Picard rank two, and it is straightforward to perform the same analysis for each. In Fig.~\ref{fig:allpicardtwo}, we show the homology regions that correspond to large black holes,  where the first quadrant always corresponds to the cone of holomorphic curves. Here the fourth and ninth examples correspond to the examples studied above. In the non-BPS case, we find that the black hole mass is always greater than the volume of the minimal piecewise-calibrated representative of the corresponding homology class. For non-holomorphic curves we find that the BPS-anti-BPS constituents always experience a repulsive force inside the K\"ahler cone. 

Examining the regions in Fig.~\ref{fig:allpicardtwo}, we see that in some cases the large non-BPS black hole region covers some holomorphic curve classes, when the blue region covers some of the first quadrants in the plots, but we still find that the mass of the black hole is greater than the mass of the minimal piecewise-calibrated representative, which itself is calibrated. Therefore, even for some holomorphic curves we predict the existence of a connected local, but not global, volume minimizer. This may come as a surprise, as a particle corresponding to an M2 brane on a holomorphic curve is a BPS state. However, when we gather many of these particles in order to form a black hole, it may be that there is actually no BPS solution to the black hole equations of motion, as is the case in the conifold. For some cases instead there is a non-BPS black hole, which is interpreted as an M2 brane wrapping a connected representative of a holomorphic curve class that itself is not a global volume minimizer, and so is not calibrated.

For examples with higher Picard rank the analysis becomes more involved, and in particular it is not always possible to obtain analytic results. However, we performed a scan over the rest of the Calabi-Yau hypersurfaces in smooth Fano toric fourfolds, considering the following possibilities for non-effective curves: 1) all tuples of the generators of the semigroup of effective curves with coefficients plus or minus one, 2) identifying BPS black holes and flipping the sign of some of the constituent curve classes, and 3) random combinations of rational curves of mixed sign. In all cases we found that the black hole mass was always greater than the volume of the minimal piecewise-calibrated representative of the corresponding homology class. We take this as suggestive evidence that if curve recombination does occur in Calabi-Yau threefolds for LBBC's, it is rare.

\subsection{Elliptic fibrations}

We will now consider smooth elliptic fibrations $\pi : X \rightarrow B$ over simple bases $B$, with a section. These examples are particularly interesting because we will be able to compare our results for the non-BPS black hole entropy with a microscopic prediction based on the BPS calculation related to oscillator modes of a string as in~\cite{Vafa:1997gr,Haghighat:2015ega}. This will be discussed in more generality in \S\ref{sec:f-theory}.

Denote the section as $\sigma$, and its dual $(1,1)$-form as $\omega_0$. The other divisors are pullbacks of curves $C_\alpha$ in the base $B$, denoted $D_\alpha = \pi^*(C_\alpha)$, and denote their dual $(1,1)$-forms as $\omega_\alpha$. We expand the K\"ahler form as $J = t_0 \omega_0 + t^\alpha \omega_\alpha$. The triple intersection numbers are~\cite{Couzens:2017way}
\begin{align}
& C_{000} = \int\limits_B c_1(B)^2 = 10 - h^{1,1}(B)\, ,\nonumber\\
& C_{00\alpha} = -c_1(B)\cdot C_\alpha\, ,\nonumber \\
& C_{0\alpha \beta} = C_\alpha \cdot C_\beta := \Omega_{\alpha \beta}\, ,
\end{align}
where $\Omega_{\alpha \beta}$ is the intersection matrix of the $C_\alpha \subset B$. The volume of $X$ then takes the form
\begin{equation}
\mathcal{V} = \frac{1}{6}\left(C_{000}t_0^3 + 3C_{00\alpha}t_0^2 t^\alpha + 3 t_0 \Omega_{\alpha \beta}t^\alpha t^\beta \right)\, .
\end{equation}
For a smooth elliptic fibration over $X\rightarrow B$, the algebraic curves are the curves inherited from the base $B$, and the class of the fiber, which is generally thought to be a pullback of an ample divisor on $B$~\cite{borcea1991homogeneous}.

\subsubsection{$B = \mathbb{P}^2$}\label{sec:P2baseBH}
Let us start with the simplest example. There is only a single generator for divisors on $\mathbb{P}^2$, and so the fiber class $E$ must be proportional to $D^2$, where $D$ is the inverse image of the hyperplane section on $\mathbb{P}^2$. The intersection of the fiber class with the section is $\sigma \cdot E = 1$, and so we can identify the class of the typical fiber as $C_1 := E = D^2$. The curve in the base is given by $ C_2 =\sigma \cdot D$. A general curve can then be written as
\begin{equation}
C = \alpha C_1 + \beta C_2\, .
\end{equation}
We have the intersection numbers
\begin{align}
&\sigma \cdot C_1 = 1\, ,\nonumber \\
&\sigma \cdot C_2 = -c_1(B)\cdot_B \pi(D) = -3\, ,\nonumber \\
&D \cdot C_1 = 0\, ,\nonumber \\
& D\cdot C_2 = 1\, ,
\end{align}
and so we can read off the charges as
\begin{equation}
q_0 = \alpha -3\beta\, , \quad q_1= \beta\, .
\end{equation}
The K\"ahler cone conditions read $t_1 - 3t_0 > 0, t_0 > 0$, and the volume takes the form
\begin{equation}
\mathcal{V} = \frac{1}{2} t_0 \left(3 t_0^2-3 t_0 t_1+t_1^2\right)\, .
\end{equation}
It is simplest to define new K\"ahler parameters $b_1 = t_1 - 3t_0$ and $b_2 = t_0$, so that the K\"ahler cone conditions read $b_1, b_2 >0$. In this basis the volume takes the form
\begin{equation}
\mathcal{V} = \frac{1}{2} b_2 \left(b_1^2+3 b_1 b_2+3 b_2^2\right)\, .
\end{equation}
The black hole potential takes the form
\begin{align}
V_{eff} = \frac{1}{b_1(b_1+3b_2)} (\beta ^2 b_1^4+2 b_1^2 b_2^2 \left(\alpha ^2-3 \alpha  \beta +9 \beta
   ^2\right)+6 \beta ^2 b_1^3 b_2\nonumber \\+ 6 b_1 b_2^3 \left(\alpha ^2-4
   \alpha  \beta +6 \beta ^2\right)+3 b_2^4 (\alpha -3 \beta )^2 )\, ,
\end{align}
subject to the constraint $\mathcal{V} = 1$.  Define the ratio of K\"ahler parameters
\begin{equation}
x = \frac{b_1}{b_2}\, .
\end{equation}
Let us first determine the conditions for a BPS black hole, which corresponds to 
\begin{equation}
\beta  (x+3)^2=\alpha  (2 x+3)\, .
\end{equation}
Solving this with $x > 0$ gives 
\begin{equation}
x= \frac{\sqrt{\alpha ^2-3 \alpha  \beta }+\alpha -3 \beta }{\beta }\, ,
\end{equation}
and enforcing that $\mathcal{V} = 1$ we find
\begin{equation}
\frac{1}{2} (x (x+3)+3) b_2^3 = 1\, .
\end{equation}
In order to have $b_1,b_2 > 0$ we need to enforce that $x>0, b_2>0$, which leads to the conditions
\begin{equation}
\{\alpha < 0\,\,\, \mathrm{ and } \, \, \, \frac{\alpha}{3} <\beta < 0\} \quad \mathrm{or} \quad \{\alpha > 0\,\,\, \mathrm{ and } \, \, \, \frac{\alpha}{3} >\beta > 0\}\, .
\end{equation}

\begin{figure}
\begin{center}
  \includegraphics{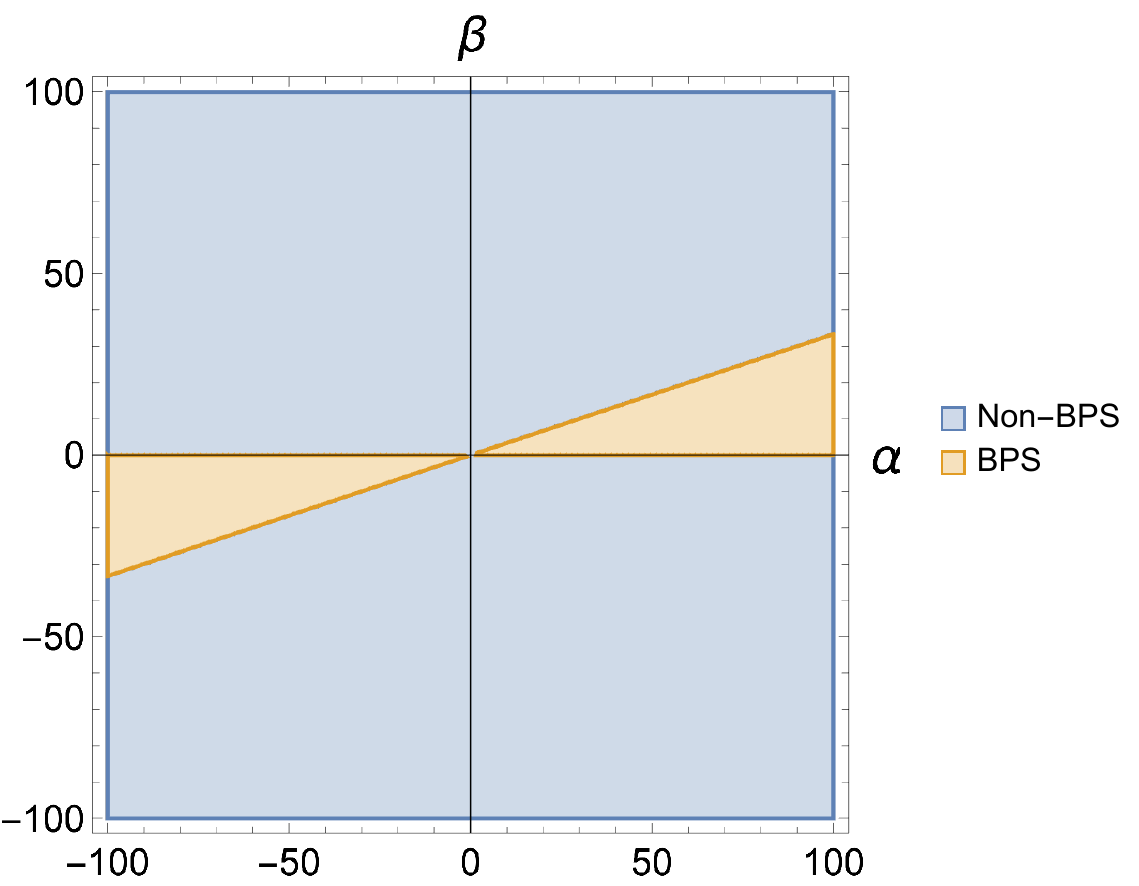}
\caption{The regions of curve space with large black holes, for a generic CY elliptic fibration over $\mathbb{P}^2$. The axes indicate the homology class, specified by $\alpha$  and $\beta$. Yellow indicates BPS black holes, while blue indicates non-BPS black holes.}\label{fig:P2elipticbh}
\end{center}
\end{figure}

Let us next determine the conditions for a non-BPS black hole, which corresponds to a solution of
\begin{equation}\label{eqn:quinticpoly2}
2 \beta  x^5 -81 \beta +x^4 (4 \alpha +9 \beta )+x^3 (24 \alpha -9 \beta )+x^2 (51 \alpha
   -90 \beta )+x (45 \alpha -135 \beta ) + 27 \alpha=0\, .
\end{equation}
Again, in order to have $b_1,b_2 > 0$ we need to enforce that $x>0, b_2>0$. Eq.~\ref{eqn:quinticpoly2} has a positive root when
\begin{equation}
\{\alpha < 3\beta\,\,\, \mathrm{ and } \, \, \, \beta > 0\} \quad \mathrm{or} \quad \{\alpha > 3\beta\,\,\, \mathrm{ and } \, \, \, \beta < 0\}\, ,
\end{equation}
which fills out the entire charge space except for codimension-one regions, which we plot in Fig.~\ref{fig:P2elipticbh}. All of the solutions are attractors.

\begin{figure}
\begin{center}
  \includegraphics{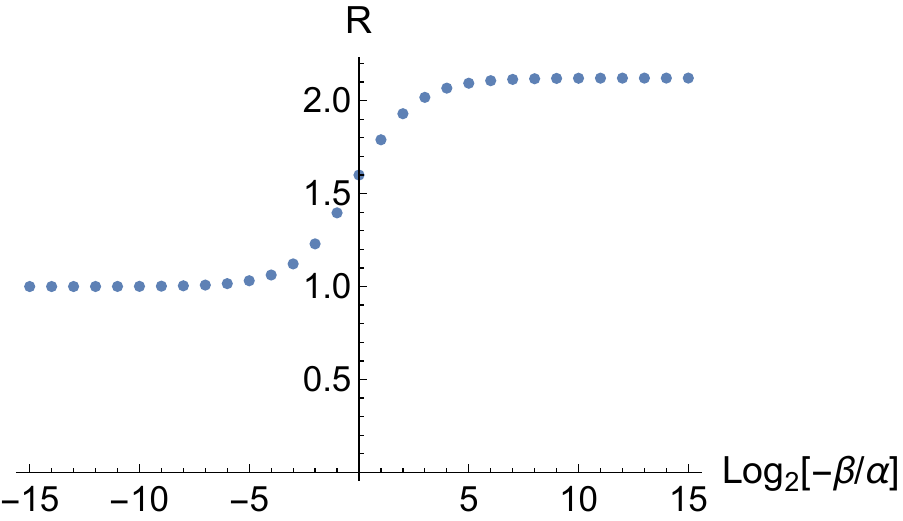}
\caption{The ratio of the black hole mass to the volume of the minimal piecewise-calibrated representative, as a function of $\beta/\alpha$, for a generic CY elliptic fibration over $\mathbb{P}^2 $. This ratio is always greater than unity, but approaches unity as the homology class becomes ``mostly'' holomorphic with $\beta \gg \alpha$. It does not approach unity for $\alpha \gg \beta$, as this limit does not correspond to a BPS black hole.}\label{fig:P2elipticbhratio}
\end{center}
\end{figure}

Let us examine the mass of the non-BPS black holes corresponding to M2 branes on non-holomorphic curves. We cannot solve Eq.~\ref{eqn:quinticpoly2} analytically, so we will instead investigate it numerically. We plot the ratio $R$ of the black hole mass to the piecewise-calibrated volume of the corresponding cycle in Fig.~\ref{fig:P2elipticbhratio}, with the asymptotic moduli set to the attractor values. We again find $R >1$ in the non-BPS case.  Therefore these black holes are unstable to decay to their BPS-anti-BPS constituents, and correspond to M2 branes wrapping connected representatives of the corresponding homology class which are local, but not global, volume minimizers. This example has an interesting feature: for $|\beta| \ll |\alpha|$, we have $R$ trending towards unity, indicating that as the curve becomes ``mostly holomorphic'', the volume approaches that of its piecewise-calibrated representative. However, for $|\alpha| \ll |\beta|$, $R$ does not approach one, indicating the corresponding representative of the curve class is not trending towards a piecewise-calibrated representative. This is explained by the observation that in the $|\beta| \ll |\alpha|$ case we presumably approach a BPS black hole, but in the $|\beta| \gg |\alpha|$ we do not. This manifests itself in the algebraic computation of the solutions to critical points of the potential.
Consider the quantized charges of the M2 brane on the curve, given by the intersection numbers:
\begin{equation}
(q_1, q_0) = (\beta, \alpha - 3\beta)\, .
\end{equation}
For a BPS black hole, the equations of motion in Eq.~\ref{eqn:bpssoln} give
\begin{equation}
\tau_I \sim q_I\, .
\end{equation}
The divisor volumes take the form
\begin{equation}
(\tau_1, \tau_0) = \left(b_2 b_1 + \frac{3b_2^2}{2} , \frac{b_1^2}{2}\right)\,.
\end{equation}
In the case the $ |\alpha| \gg |\beta|$, the BPS equations of motion then be consistently solved in the regime $b_2 \ll b_1$, and taking a small magnitude negative $\beta$ will simply be a perturbation to the large BPS black hole. However, in the case that $|\beta| \gg |\alpha|$ we cannot approach a large BPS black hole solution, since the BPS equations of motion would force a four-cycle volume to become negative. Therefore taking $|\beta| \gg |\alpha|$ does not bring us close to a large BPS black hole, which explains why $R$ does not approach unity in that limit.

The non-BPS black hole region also covers some holomorphic curves, though these curves do not correspond to BPS black holes. Here the mass of the black hole is still larger than the minimal piecewise-calibrated representative of the corresponding curve class, which itself is calibrated. Therefore even in the case of holomorphic cycles we predict the existence of a connected local, but not global, volume minimizer, whose volume is given by the black hole mass.

In this example the force between the BPS-anti-BPS constituents takes the form
\begin{equation}
P \sim -\frac{6 \alpha  \beta  b_2^2 (b_1+b_2)}{b_1}\, , 
\end{equation}
which is repulsive everywhere inside the K\"ahler cone for mixed sign $\alpha$ and $\beta$. 

It will also be interesting to consider the large $n$ limit of these non-BPS black holes, where $n$ is the wrapping number of the elliptic fiber above, corresponding to $\alpha = n$. Let us first work out the BPS case. In the large $n$ limit we have
\begin{equation}
x = \frac{2n}{\beta}\, .
\end{equation}
Solving $\mathcal{V} = 1$, we have
\begin{equation}
b_1 = \frac{2^{2/3}n^{1/3}}{\beta^{1/3}}\, , \quad b_2 = \frac{\beta^{2/3}}{2^{1/3}n^{2/3}}\, ,
\end{equation}
and so we find
\begin{equation}
V_{eff} = 3\times 2^{1/3}n^{2/3}\beta^{4/3}\, .
\end{equation}
We then find
\begin{equation}
S = \sqrt{2}\pi \sqrt{n}\beta\, .
\end{equation}
Let us now compare this to the microscopic formula for the entropy found in~\cite{Haghighat:2015ega}. For an M2 brane wrapped on the elliptic fibers $n$ times, and a curve on the base $C$, the microscopic entropy is computed as
\begin{equation}
S_{micro} = 2\pi \sqrt{\frac{n c_L}{6}}\, , 
\end{equation}
where $c_L$ is the left-moving central charge written as
\begin{equation}
c_L = 3 C\cdot C + 9 c_1(B)\cdot C + 6\, ,
\end{equation}
and $c_1(B)$ is the first Chern class of the base. For $B = \mathbb{P}^2$ we have $C = \beta C_2$, and in the large $\beta$ limit we have $c_L = 3 \beta^2 C_2 \cdot C_2 = 3\beta^2$, and so we find
\begin{equation}\label{eqn:bpsentropyP2}
S_{micro} = \sqrt{2} \pi \sqrt{n}\beta\, , 
\end{equation}
which agrees with the black hole entropy.

In the non-BPS case, the equation of motion is given in Eq.~\ref{eqn:quinticpoly2}, which in the limit $n \gg \beta$ reduces to
\begin{equation}
2\beta ^2 x^7+21 \beta ^2 x^6-8 n ^2 x^5-60 n ^2 x^4-174 n ^2 x^3-243 n ^2 x^2-189 n ^2 x-81 n ^2 = 0\, .
\end{equation}
Taking $x\sim n$, we have
\begin{equation}
2 \beta^2  x^7  - 8n^2 x^5 =0\, ,
\end{equation}
for which we consistently find
\begin{equation}
x = -\frac{2n}{\beta}\, .
\end{equation}
Enforcing $\mathcal{V} = 1$ in the large $\alpha$ limit, we then find
\begin{equation}
b_1 = 2^{2/3}\left(-\frac{n}{\beta} \right)^{1/3}\, , \quad b_2 = \left(\frac{\beta^2}{2n^2} \right)^{1/3}\, .
\end{equation}
Keeping the leading terms in $n$ in $V_{eff}$, we then find
\begin{equation}
V_{eff} = 3\times 2^{1/3}n^{2/3}\beta^{4/3}\, ,
\end{equation}
which gives an entropy of the form
\begin{equation}
S = \sqrt{2}\pi \sqrt{|n|}|\beta|\,,
\end{equation}
and so we can read off the central charge of the theory as 
\begin{equation}
c = 3\beta^2 = 3|C\cdot C|\, ,
\end{equation}
which is simply the absolute value of the analytic continuation of the leading-order BPS central charge to non-holomorphic curves. We will return to this point in greater detail in \S\ref{sec:f-theory}.

\subsubsection{$B = \mathbb{F}_n$ and curve recombination}\label{sec:bfn}
Let us now consider the case that the base of the elliptic fibration is a Hirzebruch surface $\mathbb{F}_n$, where we take $n = 0,1,2$ so that the generic $\pi: X \rightarrow B$ is smooth. We can describe $\mathbb{F}_n$ as a toric variety via a fan with rays
\begin{equation}
v_1 = \{0,1\}\, , \quad v_2 = \{n,-1\}\, , \quad v_3 = \{1,0\}\, , \quad v_4 = \{-1,0\}\, .
\end{equation}
The projective scaling weights of the corresponding toric coordinates are
$$
\begin{array}{ c  c  c  c | c }
  x_1 & x_2 & x_3 & x_4 & c_1 \\
  0 & 0 & 1 & 1 & 2 \\
  1 & 1 & 0 & n & 2+n \\
\end{array}
$$
where $c_1$ is the first Chern class. From the fan we have the intersection structure
\begin{equation}
(1,0)\cdot (0,1) = 1\, , \quad (0,1) \cdot (0,1) = 0, \quad (1,0) \cdot (1,n) = 0\, ,
\end{equation}
and so the generating set of divisor on $\mathbb{F}_n$ is $\{C_1, C_3\}$, and we can read off $C_1^2 = 0$, $C_1 \cdot C_3 = 1$, and
\begin{equation}
C_3^2 = (1,0) \cdot (1,0) \left[(1,n) - n(0,1) \right]\cdot (1,0) = -n\, .
\end{equation}
In addition, we have $c_1 = 2D_3 + (2+n)D_1$, and so we then have
\begin{align}
& c_1 \cdot C_1 = 2\, , \nonumber\\
& c_1 \cdot C_3 = -2n + (2+n) = (2-n)\, .
\end{align}
We now consider the elliptic fibration over $\mathbb{F}_n$. To determine the class of the elliptic fiber, we note that the intersection of the fiber with the section is one. Defining $D_\alpha = \pi^\ast (C_\alpha)$, we then consider
\begin{equation}
\sigma \cdot (a D_1 + b D_3)^2 = 2 ab \sigma \cdot D_1 \cdot D_3 + b^2\sigma \cdot D_3^2 = 2 ab -nb^2\, .
\end{equation}
Setting the above equation equal to one, we have
\begin{equation}
a = \frac{1 + nb^2}{2b}\, .
\end{equation}
Choosing $b = 1$, we have $a = (1 + n)/2$. 

\begin{figure}
\begin{center}
  \includegraphics{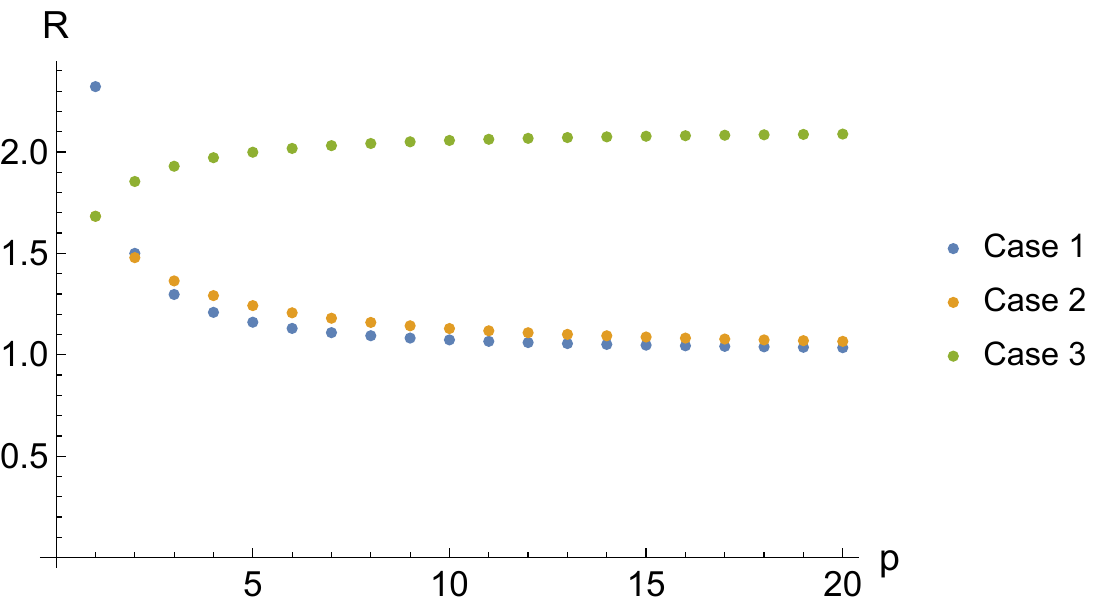}
\caption{The ratio of the black hole mass to the volume of the minimal piecewise-calibrated representative for a generic CY elliptic fibration over $\mathbb{P}^1 \times \mathbb{P}^1 $. Case 1 is the curve class $C_1 - C_3$ with $p$ wrappings of the fiber. Case 2 is the curve class $C_1 +C_3$ with $-p$ wrappings of the fiber. Case 3 is the curve class $p(C_1 + C_3)$ with a single (negative) wrapping of the fiber. This ratio is always greater than unity, but approaches unity for Case 1 and Case 2. It does not approach unity for Case 3, as this case does not approach a BPS black hole.}\label{fig:Fnelipticbhratio}
\end{center}
\end{figure}

To compute the intersection numbers we have $c_1(B) = (2,2+n)$, and so we have $C_1 \cdot c_1 = (0,1)\cdot (2,2+n) = 2$, and $C_3 \cdot c_1 = 2-n$.
The volume of $X$ then takes the form
\begin{equation}
\mathcal{V} = \frac{1}{6}\left(8 t_0^3 -6 t_0^2 t_1 - 3(2-n)t_0^2 t_3 +3t_0 (2t_1 t_3 -n t_3^2)\right)\, . 
\end{equation}
A general curve can be written $C = \alpha C_1 + \beta C_3 + \gamma E$, where $E$ is the class of the elliptic fiber. The intersection numbers with the basis of divisors is given by
\begin{align}
&D_1 \cdot C = \alpha D_1^2 \cdot \sigma + \beta D_1 \cdot D_3 \cdot \sigma = \beta\, , \nonumber \\
&D_3 \cdot C = \alpha D_3 \cdot D_1 \cdot \sigma  + \beta D_3^2 = \alpha - n\beta , , \nonumber \\
&\sigma \cdot C = \alpha \sigma^2 D_1 + \beta \sigma^2 D_2 + \gamma  = -2\alpha + (n-2)\beta + \gamma\, ,
\end{align}
and so the K\"ahler cone conditions are
\begin{align}
& t_0 > 0\, , \quad t_1 - nt_3 + (n-2)t_0 >0\, ,\quad  t_3 -2 t_0 > 0\, .
\end{align}
Let us examine some of the black hole masses for non-holomorphic curves. We focus on the $n=0$ case since it is simple and illustrative of the general behavior. We consider three case:  first, we can fix a non-holomorphic cycle in the base of the form $C_1 - C_3$, and take $p$-times the fiber (for a black hole we really need to take a large multiple of this charge, but that does not affect the analysis). Second, we fix an ample class in the base of the form $C_1 + C_3$, and take $-p$ times the fiber. Third, we can take a single wrapping of the fiber class, and take $p$ times an ample class in the base $p(C_1 + C_3) $. In Fig.~\ref{fig:Fnelipticbhratio} we plot the ratio $R$ of the black hole mass to the volume of the minimal piecewise-calibrated representative in each case, where all of the solutions correspond to attractors. For the first two cases when $p$ becomes large the black hole mass approaches the volume of a piecewise-calibrated representative of the corresponding curve class. However, for the third case $R$ does not approach unity with large $p$, indicating the corresponding representative of the curve is not approaching a piecewise-calibrated one. This is the same behavior that we observed in the elliptic fibration over $\mathbb{P}^2$, as the third limit does not approach a BPS black hole.

Let us again examine the force between the BPS and anti-BPS constituents in these cases. We again re-define our K\"ahler coordinates so that the K\"ahler cone conditions read $b_1, b_2, b_3 >0$. For almost all examples we find that this force is repulsive inside the K\"ahler cone; however, in one example we find that the force is attractive, but the black hole mass is larger than the masses of the sum of the BPS and anti-BPS constituents.  We consider the case where the M2 brane wraps (a large multiple of) the curve class $C_1 - C_3 +E$. In this case the force between the BPS-anti-BPS constituents takes the form
\begin{equation}\label{eqn:attractiveforce}
P \sim -\frac{2 b_3^2 (b_1 (b_2+b_3)+b_3 (3 b_2+4
   b_3))}{3 (b_1 (b_2+b_3)+b_2 b_3)}\, ,
\end{equation}
and is therefore attractive for all values of the moduli inside the K\"ahler cone, including the attractor background. However, we still find that the black hole mass is larger than the sum of its BPS and anti-BPS constituents. This can be resolved by considering the bound states. That is, consider an M2 brane wrapped on $C_1 +  E$, and an M2 brane wrapped on $-C_3$, separated by a small spatial distance in the non-compact 5d spacetime. In the attractor background set by the black hole, these two states are attractive, and we expect them to form a bound state that respects the attractor background, from which we can form the black hole. Let us write the mass of the bound states as
\begin{equation}
m = m_1 + m_2 - \delta\, ,
\end{equation} 
where $m_1 = \mathrm{vol}(C_1 + E)$, $m_2 = \mathrm{vol}(C_3)$, and $\delta$ is a binding energy. If we ignore the binding energy, we find that the gravitational and scalar attractive forces are large enough to overcome the electric repulsion, so these bound states seem to be mutually attractive; that is, if we collected a large cloud of the bound states they would be attracted to one another, and would collapse and release energy in the process of forming a black hole. However, we know that in this background there is an extremal black hole with the same charge, and so this appears to be a contradiction: the bound states have an attractive force, but the black hole mass is greater than the sum of the constituents. This can be rectified by including binding energy: if the bound states are tightly bound enough to lower the gravitational and scalar attraction so that the electric repulsion takes over, then one must put in energy to the system to form the black hole. This idea is consistent with the Repulsive Force Conjecture~\cite{Heidenreich:2019zkl}. This binding energy can arise in two ways: first, it can arise from a binding in the non-compact spacetime, with the particles separated by a finite distance. Second, it can arise from non-trivial recombination of the cycles $C_1 - C_3+  E$. We therefore expect this cycle to be a candidate for recombination. It was shown in~\cite{AREZZO2005209} that for $\mathbb{P}^1 \times \mathbb{P}^1$, endowed with the metric $2g_1 + 3g_2$, where $g_1,g_2$ are the standard Fubini-Study metrics on the respective projective factors, the non-holomorphic curve $C_1 - C_3$ does exhibit recombination. In our example above the base K\"ahler parameters differ by a factor of $3$ instead of $3/2$, so we take this as a suggestion of possible recombination. 

One could also consider another initial configuration from which to form the black hole. Instead of bringing together microscopic BPS and anti-BPS particles one at a time, one could gather a large cloud of the BPS particles, and another large cloud of the anti-BPS particles, and allow them to collide. However, from Eq.~\ref{eqn:attractiveforce}, we would expect the BPS and anti-BPS clouds to have an attractive force between them, and therefore lower the energy in forming the black hole! However, this does not work: in order to form a macroscopic black hole we need a large amount of charge, and the BPS and anti-BPS clouds will therefore back-react on the moduli. Near each of the clouds the black hole equations of motion force one of the moduli to become formally negative, and we therefore cannot reliably calculate the force between the clouds in this way, and cannot make sense of the black hole formation process from this initial configuration.

Again, it will also be interesting to consider the large $n$ limit of these non-BPS black holes, where $n = \gamma $ is the wrapping number of the elliptic fiber above. We will first work out a simple BPS example, where the M2 brane wraps the fiber $n$-times, and the base curve $C = C_1 + C_3$ $\beta$-times. This setup is symmetric in the base K\"ahler parameters and so we can set them to be equal. In the large $n$ limit we find
\begin{equation}
V_{eff} = 6\beta^{4/3}n^{2/3}\, ,
\end{equation}
which gives an entropy of the form
\begin{equation}\label{eqn:fnentropybps}
S = 2\pi \beta \sqrt{n}\, .
\end{equation}
Again we can compare to the microscopic entropy computed in~\cite{Haghighat:2015ega}. We have
\begin{equation}
C\cdot C = 2\beta^2\, ,
\end{equation}
and so $c_L = 6\beta^2$ in the large charge regime. The microscopic entropy is then computed as
\begin{equation}
S = 2\pi \sqrt{\frac{n c_L}{6} } = 2\pi \beta \sqrt{n}\, ,
\end{equation}
in agreement with the macroscopic entropy computed via the BPS black hole. 

For the non-BPS case, we consider an M2 brane wrapped $n$ times around the elliptic fiber, and $\beta$ times around the base curve $C_1 - C_3$. In the large $n$ limit was can again set the base K\"ahler parameters to be equal, though this is corrected by sub-leading terms in $n$. Taking $x = b_2/b_3$, where $b_2$ is the value of the base K\"ahler parameters and $b_3$ is the fiber volume, the non-BPS equations of motion in the large $n$ limit become
\begin{equation}
12 \beta ^2 x^7+84 \beta ^2 x^6-12 n ^2 x^5-60 n ^2
   x^4-116 n ^2 x^3-108 n ^2 x^2-56 n ^2 x -16 n ^2= 0\, .
\end{equation}
Taking $x \sim n$, this reduces to
\begin{equation}
12 \beta ^2 x^7-12 n ^2 x^5 = 0\, ,
\end{equation}
which is solved by $x = -n/\beta$. Enforcing $\mathcal{V} = 1$, and solving for $b_2$ and $b_3$, we find the value for the effective potential
\begin{equation}
V_{eff} = 6\beta^{4/3}n^{2/3}\, ,
\end{equation}
giving an entropy of the form
\begin{equation}\label{eqn:fnnonbpsminus}
S^{non-BPS} = 2\pi |\beta| \sqrt{|n|}\, .
\end{equation}
The central charge is again given by the absolute value of the analytic continuation of the BPS central charge to non-holomorphic curves, of the form $3 |C\cdot C|$. 

Finally, in a similar fashion we can consider an M2 brane wrapped $n$ times around the elliptic fiber, and $\beta$ times around the base curve $C_1 + C_3$, with $\beta$ and $n$ of mixed sign. In this case we can consistently set $b_1 = b_2$ due to the symmetry in the base. We find the entropy to be 
\begin{equation}\label{eqn:fnnonbpsplus}
S^{non-BPS} = 2\pi |\beta| \sqrt{|n|}\, ,
\end{equation}
which is the same as the BPS entropy to with the sign of $n$ flipped. We will explore this further in \S\ref{sec:f-theory}.

\section{Black String Examples}\label{sec:bsexamples}

We now consider 5d black strings, obtained from wrapping M5 branes on divisors in a Calabi-Yau threefold. A major difference between the black string case and the black hole case is that, in some examples, we will find that the tension of the non-BPS black string is \textit{less} than the volume of the minimal piecewise-calibrated representative of the corresponding divisor class, which we interpret geometrically as recombination. Our first two examples exhibit recombination.

\subsection{The bi-cubic in $\mathbb{P}^2 \times \mathbb{P}^2$}
We again consider a generic anticanonical hypersurface $X \subset \mathbb{P}^2 \times \mathbb{P}^2 := V$, as in Sec.~\ref{sec:p2p2bh}.
The effective cone of divisors of a generic anti-canonical hypersurface in $\mathbb{P}^2 \times \mathbb{P}^2$ is generated by $D_1$ and $D_2$~\cite{2013arXiv1305.0537O}. The black string effective potential can be written as
\begin{equation}
V_{eff} = \frac{9}{2} \left(p_1^2 t_2^2 \left(2 t_1^2+2 t_1 t_2+t_2^2\right)+2 p_1 p_2 t_1^2
   t_2^2+p_2^2 t_1^2 \left(t_1^2+2 t_1 t_2+2 t_2^2\right)\right)\, ,
\end{equation}
subject to the constraint $\mathcal{V} = 1$.  Again define the ratio of K\"ahler moduli
\begin{equation}
x = \frac{t_1}{t_2}\, .
\end{equation}
Let us first determine the conditions for a BPS black string, which corresponds to 
\begin{equation}
x = \frac{p_1}{p_2}\, . 
\end{equation}
Clearly in order to have $t_1,t_2 > 0$ we need to enforce that $x>0, t_2>0$, which leads to the conditions
\begin{equation}
\{p_1 < 0\,\,\, \mathrm{ and } \, \, \, p_2 < 0\} \quad \mathrm{or} \quad \{p_1 > 0\,\,\, \mathrm{ and } \, \, \, p_2 > 0\}\, .
\end{equation}

Let us next determine the conditions for a non-supersymmetric black string, whose equation is given by
\begin{equation}\label{eqn:quinticpoly}
 2 p_2 x^4 + x^3 (2 p_1+5 p_2)+x^2 (3 p_1+3 p_2)+x (5 p_1+2 p_2)+2 p_1=0\, .
\end{equation}
Again, in order to have $t_1,t_2 > 0$ we need to enforce that $x>0, t_2>0$. Eq.~\ref{eqn:quinticpoly} has a positive root in the case that $p_1$ and $p_2$ are of mixed sign, and so the non-BPS black string region is the complement of the BPS black string region (except for $p_1 = 0$ or $p_2 = 0$):
\begin{equation}
\{p_1 < 0\,\,\, \mathrm{ and } \, \, \, p_2> 0\} \quad \mathrm{or} \quad \{p_1> 0\,\,\, \mathrm{ and } \, \, \, p_2< 0\}\, .
\end{equation}
It can be explicitly checked that all solutions are attractors.

\begin{figure}
\begin{center}
  \includegraphics{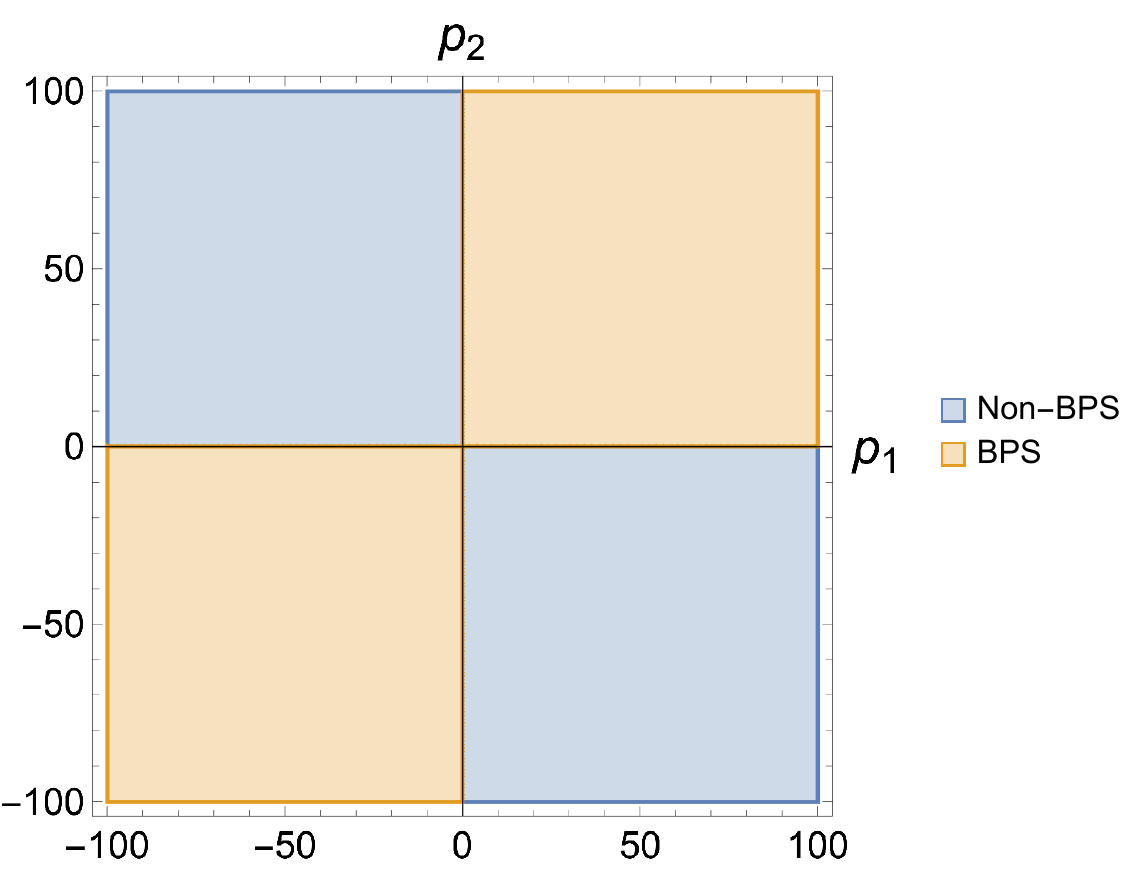}
\caption{The regions of divisor space with large black strings, for a generic CY hypersurface in $\mathbb{P}^2 \times \mathbb{P}^2$. The axes indicate the homology class, specified by $p_1$  and $p_2$. All divisors, except $p_1 = 0$ or $p_2 = 0$, correspond to large black strings.}\label{fig:P2P2bs}
\end{center}
\end{figure}

In Fig.~\ref{fig:P2P2bs} we show the regions where large black strings exist for both the BPS and non-BPS cases. In this example all divisor classes correspond to large black strings, except when $p_1 =0$ or $p_2 = 0$.

\begin{figure}
\begin{center}
  \includegraphics[scale = 1.3]{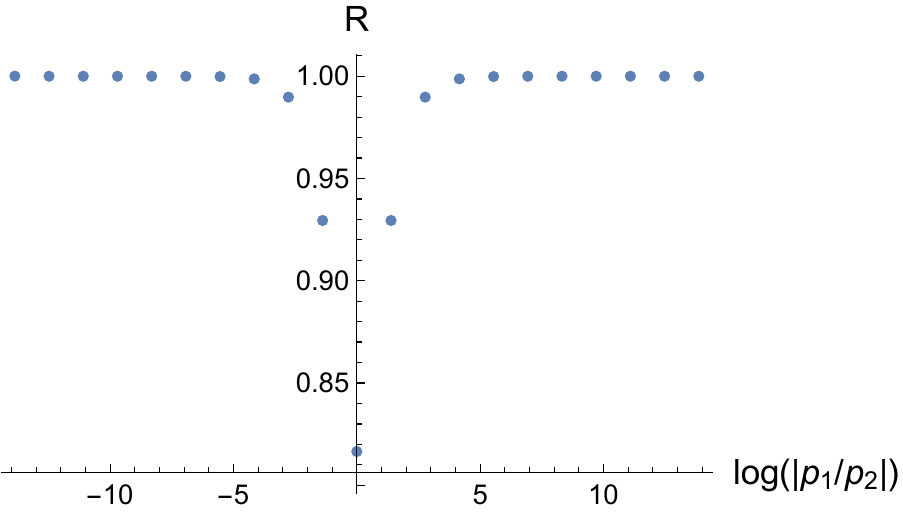}
  \caption{The ratio of the black string tension to the volume of the minimal piecewise-calibrated representative, as a function of $p_2/p_1$, for negative $p_2$ and positive $p_1$. This ratio is always less than one, indicating recombination, and that the non-BPS black string is stable against complete decay into BPS-anti-BPS constituents.}
  \label{fig:unstableP2P2mag}
  \end{center}
\end{figure}

The non-BPS equation of motion can be solved analytically, but the expression is long, and so for the non-BPS case we can simply numerically minimize the effective potential as a function of $p_2/p_1$. In this example we find that the black string tension is less than the volume of a piecewise-calibrated representative, indicating that the non-BPS black string is stable against complete decay into BPS-anti-BPS constituents, and that the corresponding non-holomorphic divisor class undergoes recombination. In Fig.~\ref{fig:unstableP2P2mag} we plot the ratio $R$ of the black string tension the volume of the piecewise-calibrated representative of the corresponding divisor class, against the ratio $p_2/p_1$, for negative $p_2$ and positive $p_1$. Note that, as one of the charges becomes much larger in magnitude than the other, the ratio approaches unity, suggesting that the volume minimizing representative may approach an (anti)-holomorphic representative, as the solution approaches a BPS one.

Without loss of generality, choose $p_1 >0$ and $p_2 <0$, and as usual fix the asymptotic moduli to the attractor values. In general, the non-BPS string tension predicts that the non-holomorphic homology class $p_1 D_1 + p_2 D_2$ has minimal volume less than that of its minimal piecewise-calibrated representative, indicating that non-trivial recombination has occurred. This example also makes a non-trivial geometric prediction for the WGC, namely that this black string should be able to decay. A particularly instructive configuration is the black string with $p_1 = -p_2 = n$, with $n \gg 1$. In this case we find that the ratio of the black string tension $T$ to the minimal piecewise-calibrated volume $T^\cup$ is
\begin{equation}
\frac{T}{T^\cup} = \sqrt{\frac{2}{3}}\, .
\end{equation}
We can also ask what decay channels are allowed for this black string. The simplest possibility is for the black string to emit a BPS (or anti BPS) string, for instance taking $p_1 \rightarrow p_1 - 1$. The black string tension before the emission is given by
\begin{equation}
T_b = \sqrt{2} 3^{5/6} n\, ,
\end{equation}
while the tension after emission is given by\footnote{We take $n$ to be large, in which case the change in horizon moduli under the shift in black string charge is negligible.}
\begin{equation}
T_a = \sqrt{2} 3^{5/6} \sqrt{n (n-1 )}\, .
\end{equation}
Taking the difference in the large-$n$ limit, we have
\begin{equation}
\delta T = \frac{3^{5/6}}{\sqrt{2}}+ \mathcal{O}\left(\frac{1}{n} \right)\, .
\end{equation}
In this background, the tension of the BPS string with $p_1 = 1$ is given by
\begin{equation}
T_{BPS} = \frac{3^{4/3}}{2}\, ,
\end{equation}
and so we have
\begin{equation}
\delta T \approx 1.77 < T_{BPS} \approx 2.16\, .
\end{equation}
Therefore the decay channel of emitting a BPS string, corresponding to a generating divisor of the effective cone, is not allowed.  On the other hand the WGC suggests that it will decay to some strings carrying small (microscopic) charges. 
Since the black string has charge $(n,-n)$ the simplest possibility is that it will be able to decay by emitting a non-BPS strings of charge $(1,-1)$, or some small multiple. Let us bound the tension of this string. The tension of the black string after emission is given by
\begin{equation}
T_a = \sqrt{2} 3^{5/6} (n-1)\, ,
\end{equation}
and so we find
\begin{equation}\label{eqn:deltat}
\delta T = \sqrt{2} 3^{5/6}\, ,
\end{equation}
which gives an upper bound for the tension of the emitted string. Note that it is also possible that recombination does not occur but rather the string with charge $(1,-1)$ is composed of two disconnected BPS-anti-BPS strings which form a bound state in the 5d non-compact spacetime, lowering its energy. Since the minimal piecewise-calibrated volume of that cycle is $3^{4/3}$, if $D_1$ and $-D_2$ do recombine, via Eq.~\ref{eqn:deltat} the black string physics predicts a volume reduction of
\begin{equation}
\frac{T}{T^\cup} = \sqrt{\frac{2}{3}} \simeq 0.82\, .
\end{equation}

The force between the strings with mixed sign $p_1$ and $p_2$ is attractive inside the K\"ahler cone, and goes as
\begin{equation}
P \sim \frac{p_1 p_2}{(t_1 + t_2)^2}\, .
\end{equation}

\subsection{The tetraquadric}
In this section we consider the tetraquadric, which is a generic anti-canonical hypersurface in $(\mathbb{P}^1)^4$ of multi-degree $(2,2,2,2)$. Here we have $h^{1,1}(X) = 4$, and the Mori cone is inherited from that of the ambient product of projective spaces, which is smooth Fano. Expand the K\"ahler form in terms of duals to the hyperplane section restricted to $X$ as $J = t^i \omega_i$, with $t^i > 0$ inside the K\"ahler cone. The volume takes the form
\begin{equation}
\mathcal{V} = 2 t_1 t_2 t_3 + 2 t_1 t_2 t_4 + 2 t_1 t_3 t_4 + 2 t_2 t_3 t_4\, .
\end{equation}
We will focus on a particularly simple example, which is that of an M5 brane on the non-holomorphic divisor class $\{n,n,-n,-n\}$, where the vector indicates the wrapping numbers around the restrictions of the four hyperplanes to $X$. In this case the moduli are stabilized at the symmetric attractor point $t_I = 1/2$, and the effective potential takes the form
\begin{equation}
V_{eff} = 16 n^2\, ,
\end{equation} 
and so the black string tension, and prediction for the minimal volume of the homology class labeled by $\{n,n,-n,-n\}$, is
\begin{equation}
T_{bs} = \mathrm{vol}(\{n,n,-n,-n\}) = \sqrt{\frac{3}{2}V_{eff}} = 2\sqrt{6}|n|\, .
\end{equation}
To determine whether this non-holomorphic representative is minimal in its class, we need to identify the smallest piecewise-calibrated representative in that class. There is a canonical piecewise-calibrated representative, whose volume is given by summing the volumes of each (anti)-holomorphic hyperplane constituent of $\{n,n,-n,-n\}$, which is given by
\begin{equation}
\mathrm{vol}^\cup (\{n,n,-n,-n\}) = |n|\tau_1 + |n|\tau_2 + |n|\tau_3 +|n|\tau_4\, ,
\end{equation}
where $\tau_I$ is the volume of the $I$-th hyperplane class. Since for this set of charges the moduli flow to a symmetric point $t_I = 1/2$, we can write $\tau_I \equiv \tau = 3/2$, and we then have
\begin{equation}
\mathrm{vol}^\cup (\{n,n,-n,-n\}) = 4|n|\tau = 6|n|\, . 
\end{equation}
Since $2\sqrt{6} \simeq 4.6<6$, this example is a candidate for exhibiting recombination. However, the effective cone of $X$ is infinitely generated~\cite{Constantin:2018hvl}, and more care is required. The semigroup of effective divisors is generated by the hyperplanes, as well as permutations of the line bundle with multi-degree $\{-1,1,1,1 \}$~\cite{Constantin:2018hvl}.\footnote{These line bundles are not themselves effective~\cite{Constantin:2018hvl}, but when combined with the hyperplanes they generate all effective line bundles.} The cone $\mathcal{C}$ generated by these line bundles therefore contains the effective cone of $X$, and all effective divisors on $X$ can be expressed as linear combinations of the eight generators of $\mathcal{C}$. We will look for the smallest piecewise-calibrated representative of the class $\{n,n,-n,-n\}$ using the generators of $\mathcal{C}$, which are
\begin{align}
& \mathcal{L}_1 = \{1,0,0,0\}\, , \nonumber \\
& \mathcal{L}_2 = \{0,1,0,0\}\, , \nonumber \\
& \mathcal{L}_3 = \{0,0,1,0\}\, , \nonumber \\
& \mathcal{L}_4 = \{0,0,0,1\}\, , \nonumber \\
& \mathcal{L}_5 = \{-1,1,1,1\}\, , \nonumber \\
& \mathcal{L}_6 = \{1,-1,1,1\}\, , \nonumber \\
& \mathcal{L}_7 = \{1,1,-1,1\}\, , \nonumber \\
& \mathcal{L}_8 = \{1,1,1,-1\}\, .
\end{align}
Consider the corresponding divisors dual to these line bundles. At the symmetric locus $t_Is = 1/2$, the ``volumes'' of the generators of $\mathcal{C}$ are $3/2$ for $i = 1\dots 4$, and are 3 for $i = 5\dots 8$. The volume of any piecewise-calibrated representative of the class $\{n,n,-n,-n\}$ can then be written as 
\begin{equation}
\mathrm{vol}^\cup =\frac{3}{2}\left(|a_1| + |a_2| + |a_3| + |a_4| + 2 |a_5| + 2|a_6| + 2|a_7| + 2|a_8| \right)\, ,
\end{equation}
where the $a_i, i = 1\dots 8$, satisfy
\begin{align}
&a_1 -a_5 + a_6 + a_7 + a_8 = n\,, \nonumber \\ 
&a_2 +a_5- a_6 + a_7 + a_8 = n\,, \nonumber \\ 
&a_3 +a_5 + a_6 - a_7 + a_8 = -n\,, \nonumber \\ 
&a_4 +a_5 + a_6 + a_7 - a_8 = -n\, .
\end{align}
Solving these equations to eliminate $a_5,a_6,a_7,a_8$, we are then left with the problem of minimizing the function
\begin{align}\label{eqn:tomin}
\mathrm{vol}^\cup = \frac{3}{4} (| a_1-a_2-a_3-a_4-2 n| + \nonumber \\
|a_1+a_2+a_3-a_4-2 n| + \nonumber \\|
   a_1+a_2-a_3+a_4-2 n| +\nonumber \\
   |a_1-a_2+a_3+a_4+2 n| +\nonumber \\
   2 | a_1| +2 | a_2| +2 |a_3| +2 | a_4| )\, .
\end{align}
The $a_i$ are bounded to the domain $\{ -4n, 4n\}$, as if the values were larger or smaller, $\mathrm{vol}^\cup$ would be larger than the candidate minimum of $6|n|$ derived above. The objective function $\mathrm{vol}^\cup$ is linear away from the hyperplane loci defined by the vanishing of the various absolute values in Eq.~\ref{eqn:tomin}, and so we are left with a linear optimization problem with a boundary. It is well known that the extrema are located at the vertices of the boundary, and so we need only to check a finite number of points to find the minimal piecewise-calibrated representative. We find a minimal value of $6|n|$, confirming that the piecewise-calibrated representative constructed from the hyperplane sections gives the minimal piecewise-calibrated volume, and thus the minimal volume of the class $\{n,n,-n,-n\}$ given by the black string tension is smaller than the volume of any piecewise-calibrated representative. This again is indicative of the presence of a stable non-BPS string.

One can check that for the above wrapping numbers, the force between the BPS and anti-BPS constituents is attractive inside the K\"ahler cone:
\begin{equation}
P \sim -\frac{n^2(t_1^2 + t_2^2)((t_3^2 + t_4^2)}{t_1 t_2 t_3 + t_1 t_2 t_4 +t_1 t_3 t_4 + t_2 t_3 t_4 )^2}\, .
\end{equation}

\subsection{A K3 fibration}

This example includes non-effective divisor classes that do not exhibit recombination, which differs from the previous two examples. We consider a generic anticanonical hypersurface $X \subset \mathbb{P}^3 \times \mathbb{P}^1$, with corresponding hyperplane divisor classes $D_1$ and $D_2$ (restricted to the hypersurface), as in \S\ref{sec:k3bh}. 

A basis of effective divisors is given by $\{D_1, D_2 \}$. However, in order to compare our the tension of our black string to a piecewise-calibrated representative of the corresponding homology class, we need to determine the effective cone of $X$. We utilize the results of~\cite{2013arXiv1305.0537O}, which give the effective cone of hypersurfaces in products of projective spaces. In particular, as we observed in the previous example, when there is a $\mathbb{P}^1$ factor in the ambient space one often finds effective divisors on the hypersurface that are not inherited from effective divisors on the ambient space (referred to as ``\textit{autochthonous divisors}'' in~\cite{Demirtas:2018akl}). In this example, the effective cone is generated (over $\mathbb{R}$) by the divisors $D_2$ and $D_3 := 4D_1 - D_2$. $D_1,D_2,D_3$ all correspond to integral divisors on $X$ themselves.

The black string effective potential can be written as
\begin{equation}
V_{eff} = 16 p_1^2 t_1^2 t_2^2+\frac{16}{3} p_1^2 t_1^3
   t_2+\frac{2 p_1^2 t_1^4}{3}+\frac{8}{3} p_1 p_2
   t_1^4+8 p_2^2 t_1^4\, ,
\end{equation}
subject to the constraint $\mathcal{V} = 1$.  Again define
\begin{equation}
x = \frac{t_1}{t_2}\, .
\end{equation}
The BPS equations of motion give
\begin{equation}
x = \frac{p_1}{p_2}\, . 
\end{equation}
Clearly in order to have $t_1,t_2 > 0$ we need to enforce that $x>0, t_2>0$, which leads to the conditions
\begin{equation}
\{p_1 < 0\,\,\, \mathrm{ and } \, \, \, p_2 < 0\} \quad \mathrm{or} \quad \{p_1 > 0\,\,\, \mathrm{ and } \, \, \, p_2 > 0\}\, .
\end{equation}

Let us next determine the conditions for a non-supersymmetric black string, whose equation is given by
\begin{equation}\label{eqn:quinticpoly5}
p_1 (x+3)+3 p_2 x=0\, ,
\end{equation}
for which the solution is
\begin{equation}
x = -\frac{3 p_1}{p_1 + 3p_2}\, .
\end{equation}
Again, in order to have $t_1,t_2 > 0$ we need to enforce that $x>0, t_2>0$, which gives the conditions
\begin{equation}
\{p_1 < 0\,\,\, \mathrm{ and } \, \, \, p_2> -\frac{p_1}{3}\} \quad \mathrm{or} \quad \{p_1> 0\,\,\, \mathrm{ and } \, \, \, p_2< -\frac{p_1}{3}\}\, .
\end{equation}
These regions of non-BPS solutions all correspond to attractors.

\begin{figure}
\begin{center}
  \includegraphics{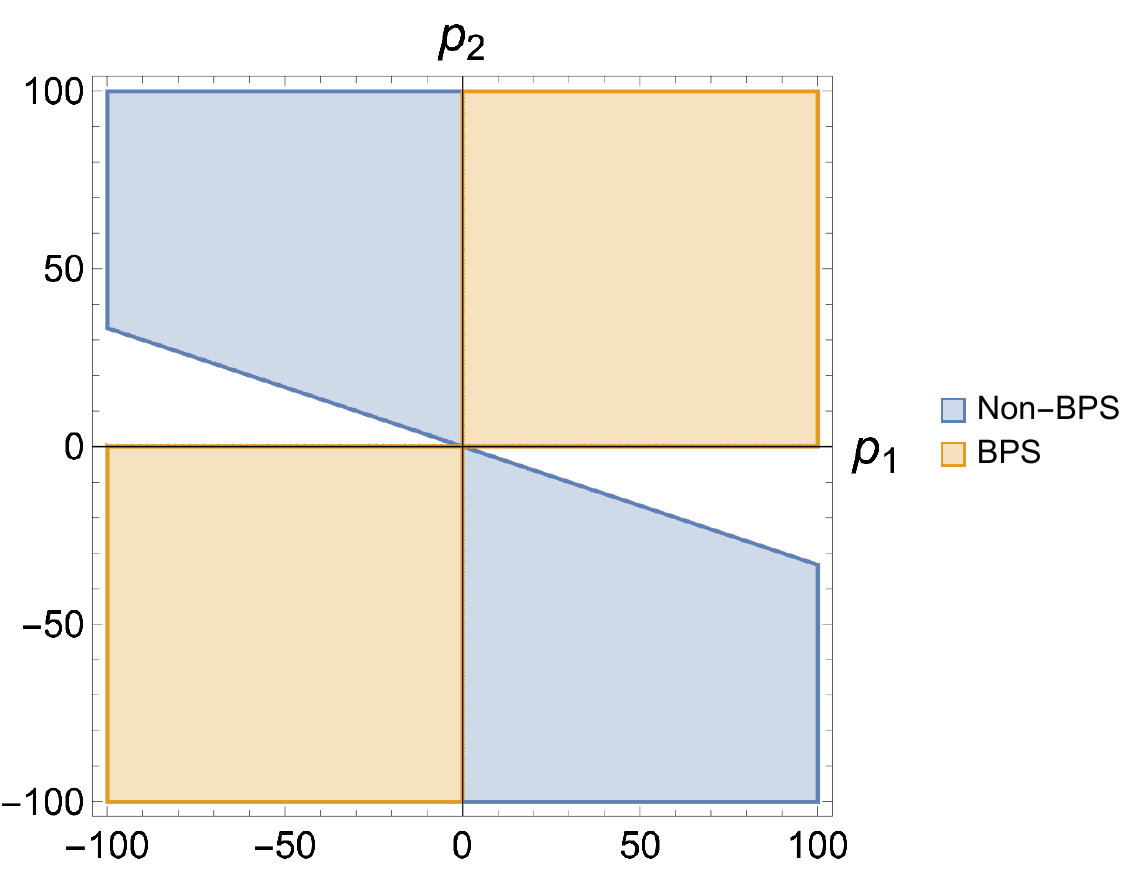}
\caption{The regions of divisor space with large black strings, for a generic CY hypersurface in $\mathbb{P}^3 \times \mathbb{P}^1$. The axes indicate the homology class, specified by $p_1$  and $p_2$. The white region indicates divisor classes that do not correspond to large black strings.}\label{fig:P3P1bs}
\end{center}
\end{figure}

In Fig.~\ref{fig:P3P1bs} we show the regions where large black strings exist for both the BPS and non-BPS cases. There are some regions that do not correspond to large black strings, including the ray along the ``extra'' generator of the effective cone $4 D_1 - D_2$.

\begin{figure}
\begin{center}
  \includegraphics{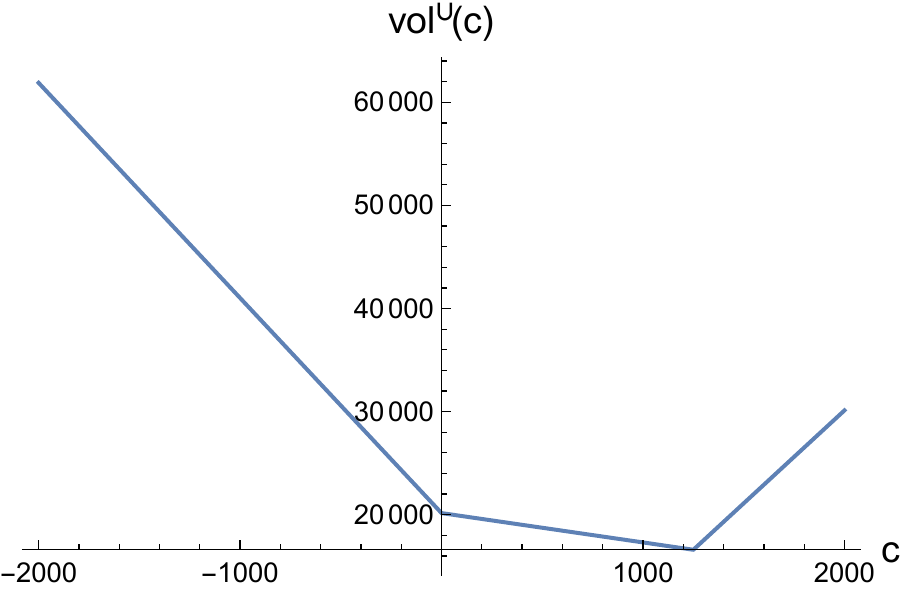}
\caption{The volume of the piecewise-calibrated representative as a function of the parameter $c$, with $n = 5000$. At large charges the presence of the extra effective divisor allows the black string to decay in BPS-anti-BPS constituents.}\label{fig:piecewise}
\end{center}
\end{figure}

We check for recombination by computing the tension of the black strings, and comparing to the piecewise-calibrated representatives of the corresponding class. Here we have three generators of integral effective divisors. Since this example has Picard rank two the effective cone is simplicial, but we need to be careful that we only consider piecewise-calibrated representatives that correspond to integral classes. Take, for example, $p_1 = n$, $p_2 = -n$, for $n \gg 1$. In order to find the volume-minimizing piecewise-calibrated representative of $n(D_1 - D_2)$, we need to consider a general family of piecewise-calibrated representatives
\begin{equation}
a D_1 + b D_2 + c D_3\, ,
\end{equation}
with
\begin{equation}
a + 4c = p_1 = p\, \quad b + c =p_2 =  -p\, ,
\end{equation}
from which we can solve for $a$ and $b$, leaving $c$ as a free parameter parameterizing the choice of piecewise-calibrated representative.
We find that the black string tension is $3.56 n$. Taking e.g., $n = 5000$, the black string tension is then $17784.5$. To compare to the minimal piecewise-calibrated representative in that class we minimize the family of piecewise-calibrated representatives with respect to $a,b,c$, demanding that $c$ is an integer. We find the minimum at $c = 1250$ (see Fig.~\ref{fig:piecewise} for the volume of the piecewise-calibrated representative $\mathrm{vol}^\cup(c)$ as a function of $c$), and the minimal piecewise-calibrated volume to be $16598.8$, less than the tension of the black string. This black string is then unstable to decay into BPS-anti-BPS constituents. In general, for large charges we find that the black string tension is greater than the volume of the minimal piecewise-calibrated representative of the corresponding divisor class. However, let us apply the black string tension formula for small charges, namely $p_1 = -p_2 = 1$, even though we do not expect it to be valid. In this case, we find that the black string tension is $3.56$, but integrality of the piecewise-calibrated representative then gives that the minimal piecewise-calibrated volume is $4.03$. Thus, for large charges, the ``extra'' effective divisor enters to allow for decay into BPS-anti-BPS constituents. This is possible suggestive evidence that the divisor $D_1 - D_2$ recombines.

 In general, decomposing the black string charge into its minimal BPS and anti-BPS constituents, we always find a repulsive force between them. For large charges the force takes the form
\begin{equation}
P \sim -\frac{3p_1 (p_1 + 4 p_2)}{8(t_1 + 6 t_2)^2}\, ,
\end{equation}
which is repulsive in the charge region where large non-BPS black strings exist for all values of K\"ahler moduli.

\subsection{Elliptic fibrations}

We consider the same elliptic fibrations for M5 branes on divisors as we did for M2 branes on curves.

\subsubsection{$B = \mathbb{P}^2$}
The generating divisor classes are the section (base), and the inverse image of the hyperplane section in the base $\mathbb{P}^2$. Denote the homology class of the M5 brane as $(p_0, p_1)$, where $p_0$ is the wrapping number around the base, and $p_1$ is the wrapping number around the vertical divisor. The effective potential is given by
\begin{equation}
V_{eff} = b_1^2 b_2^2 \left(9 p_0^2-3 p_0 p_1+p_1^2\right)+3
   b_1^3 b_2 p_0^2+\frac{b_1^4 p_0^2}{2}+3 b_1
   b_2^3 \left(3 p_0^2-2 p_0 p_1+p_1^2\right)+\frac{3
   b_2^4 p_1^2}{2}\, .
\end{equation}
We first consider BPS black strings. The BPS equations of motion yield
\begin{equation}
x = \frac{-3 p_0 + p_1}{p_0}\, ,
\end{equation}
and so BPS solutions exist for 
\begin{equation}
\{p_1 < 0\,\,\, \mathrm{ and } \, \, \, \frac{p_1}{3} < p_0 <0 \} \quad \mathrm{or} \quad \{p_1 > 0\,\,\, \mathrm{ and } \, \, \, \frac{p_1}{3} > p_0 > 0\}\, .
\end{equation}

Let us now consider non-BPS black strings. The non-BPS equations of motion give
\begin{equation}
27 p_0 - 9 p_1 + (36 p_0 + 3 p_1) x + (18 p_0 + 9 p_1) x^2 + (9 p_0 + 
     2 p_1) x^3 + 2 p_0 x^4 = 0\, .
\end{equation}
This example has the interesting feature that for the BPS black string charge range there also exist non-BPS black string critical points corresponding to the same homology class as the BPS ones. However, these solutions are not minima of the effective potential, but instead maxima.

\begin{figure}
\begin{center}
  \includegraphics{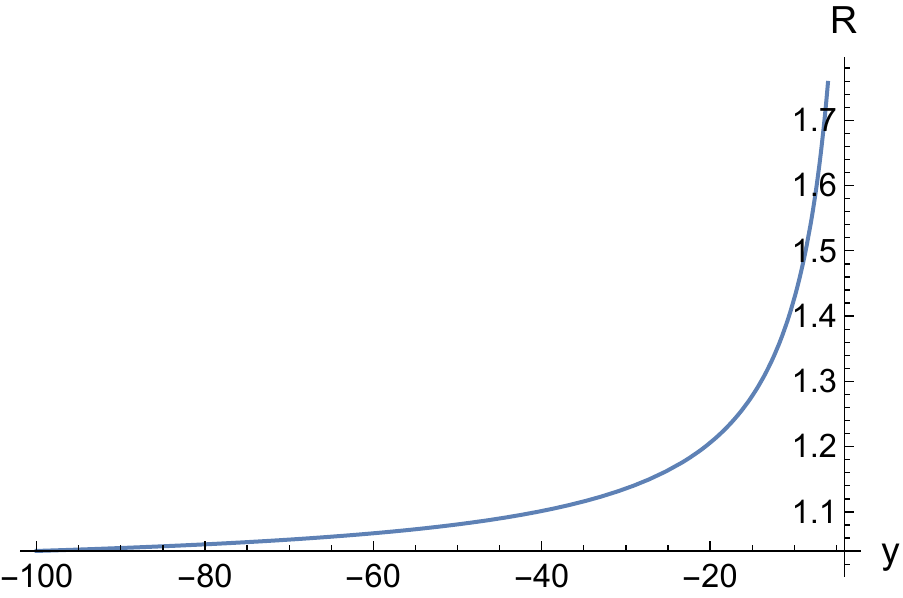}
\caption{The ratio $R$ of the black string tension to the minimal piecewise-calibrated volume of the corresponding homology class, as a function of $y = p_1/p_0$, for mixed sign $p_0$ and $p_1$. We find $R > 1$ for the entire allowed range, indicating the local representative corresponding to the black string has larger volume than the minimal piecewise-calibrated representative.}\label{fig:P2Magelipticbh}
\end{center}
\end{figure}

For mixed sign $p_0, p_1$ we find numerically there is a solution when $|p_1| \gtrsim 6 |p_0|$, and these solutions are attractors. Defining $y = p_1/p_0$, we plot $R$ in Fig.~\ref{fig:P2Magelipticbh} for such black strings. We find that $R > 1$, indicating the minimal piecewise-calibrated representative has smaller volume than the representative corresponding to the black string.

We can compute the force between the BPS and anti-BPS constituent strings, which we find to be
\begin{equation}
P \sim  -\frac{3 b_1 p_0 p_1 (b_1+2 b_2)}{2 \left(b_1^2+3
   b_1 b_2+3 b_2^2\right)^2}\, ,
\end{equation}
which is repulsive for mixed sign $p_0, p_1$ everywhere inside the K\"ahler cone. 

\subsubsection{$B = \mathbb{P}^1 \times \mathbb{P}^1$}
\begin{figure}
\begin{center}
  \includegraphics{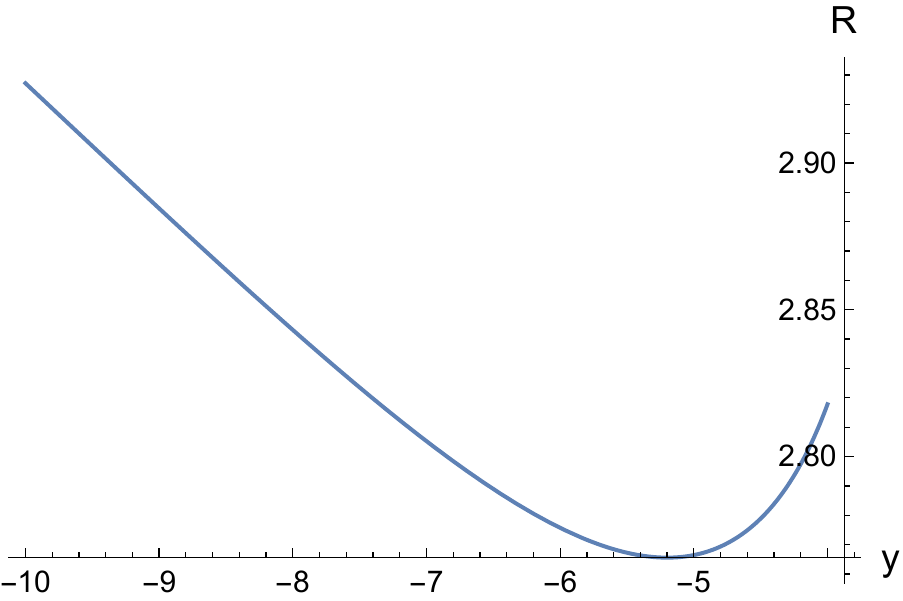}
\caption{The ratio $R$ of the black string tension to the minimal piecewise-calibrated volume of the corresponding homology class, as a function of $y = p_3/p_0$, for mixed sign $p_0$ and $p_3$. We find $R > 1$ for the entire allowed range, indicating the local representative corresponding to the black string has larger volume than the minimal piecewise-calibrated representative.}\label{fig:P1P1Magelipticbh}
\end{center}
\end{figure}

As a final black string example we will consider the base $B = \mathbb{P}^1 \times \mathbb{P}^1$, as in Sec.~\ref{sec:bfn}. There are now three divisor classes: the section, and the two vertical divisors corresponding to the inverse image of the two hyperplane curves in the base. Again let the wrapping number of the base be $p_0$, and the wrapping numbers of the vertical divisors be $p_1$ and $p_3$. We will study two interesting sub-loci in charge space. Let us first consider $p_1 = p_3$, so we can take $b_1 = b_2$ due to the symmetry in the base. Defining $x = b_2/b_3$, the BPS equations of motion give
\begin{equation}
x = \frac{-2p_0 + p_3}{p_0}\, , 
\end{equation}
and so large BPS black strings exist when 
\begin{equation}
\{p_0 > 0\,\,\, \mathrm{ and } \, \, \, p_3 > 2 p_0 \} \quad \mathrm{or} \quad \{p_0 < 0\,\,\, \mathrm{ and } \, \, \, p_3 < 2 p_0\}\, .
\end{equation}
In the case of a non-BPS black string, the equations of motion give
\begin{equation}
3 p_0 x^4 + x^3 (9 p_0+3 p_3)+x^2 (12 p_0+9 p_3)+x (16 p_0+2
  p_3)+8 p_0-4 p_3 = 0\, .
\end{equation}
We again find that for the entire BPS black string charge range there also exist non-BPS black string critical points, but again these solutions are not minima, but instead are maxima. Numerically we find that a non-BPS black string with mixed sign $p_0$ and $p_3$ exists when $|p_3|\gtrsim 4p_0$, and that these solutions are attractors. Defining $y = p_3/p_0$, we plot $R$ in Fig.~\ref{fig:P1P1Magelipticbh} as a function of $y$. We find $R > 1$, indicating the local volume minimizing representative has greater volume than the smallest piecewise-calibrated representative. 

We find that the force between the BPS-anti-BPS constituents takes the form
\begin{equation}
P \sim -\frac{3 p_0 p_3 \left(3 \left(b_1^2+b_2^2\right)+4
   b_3 (b_1+b_2)\right)}{2 (3 b_1
   (b_2+b_3)+b_3 (3 b_2+4 b_3))^2}\, ,
\end{equation}
which is repulsive for mixed sign $p_0, p_3$. 

Finally, let us consider the configuration given by $p_1 = -p_3$. In this case we will consider the limit $|p_3| \gg |p_0|$, in which case we can take $b_1 \simeq b_2$. Again we find $R > 1$, and the force between the BPS-anti-BPS constituents is repulsive.

\section{Non-BPS Black Holes from BPS and Non-BPS Strings}\label{sec:f-theory}
We have so far explored non-BPS black holes and black strings in five dimensions.  An equally interesting question is the study of black strings in 6 dimensions.
Indeed upon compactification on a circle this can lead to both black holes and black strings in 5d.  In particular here we would like to show how the non-BPS black holes in 5d can arise from BPS and non-BPS black strings in 6d and how we can use this to make a prediction for their entropy.

Consider 6d F-theory, obtained by compactification of IIb on a positively-curved space $B$, which is the base of a Calabi-Yau threefold $X$. Strings in six dimensions arise from wrapping a D3 brane on a curve $C \subset B$, and appropriate curves can lead to large extremal black strings in six dimensions, both BPS and non-BPS, as we will shortly see. The near-horizon geometry of the black strings is of the form $AdS_3 \times S^3$~\cite{Andrianopoli:2007kz}, and so is associated with a 2d CFT, regardless of whether the string is supersymmetric or not. The central charge of the CFT can be computed by the Brown-Henneaux formalism~\cite{cmp/1104114999}, and in the case of BPS strings has been verified to match the microscopic central charge~\cite{Vafa:1997gr,Maldacena:1997de}. We will extend this to the non-BPS case, providing a prediction for the central charge of the theory on the non-BPS strings. We will show that it is simply given absolute value of the analytic continuation of the BPS case: $c = 3 |C\cdot C|$.

Such black strings lead to black holes in 5d upon a circle reduction, with $n$ units of momentum around the circle. However, one can arrive at a 5d black hole with the same charges via another route, by considering M-theory on $X$, with M2 branes wrapped on the same curve $C \subset B \subset X$, and wrapping the elliptic fiber $n$ times. For a BPS black hole, where $C$ is holomorphic and $n$ is positive, this corresponds to a BPS string on a circle with $n$ units left-moving momentum on the string (reversing the orientation of the entire system also corresponds to a BPS string). In this case the macroscopic black hole entropy agrees with the Cardy formula for the leading contribution from string oscillation modes.

However, one can also consider non-BPS black holes in 5d. The mildest way to do so is to take a BPS string in 6d, compactify on a circle, but with the ``wrong'' sign of momentum which corresponds to replacing the elliptic fiber class with its conjugate $n\rightarrow -n$. For a string corresponding to a D3 brane wrapped on a holomorphic curve $C$, this corresponds to giving right-moving momentum around the circle instead of left-moving. In this case this is a non-BPS excitation of a BPS string, and we find the black hole entropy again matches the Cardy formula.

Instead, one can also take a non-BPS string in 6d and compactify on a circle. In this case we will find that the central charge from the 6d $AdS_3$ agrees with the central from the black hole entropy, when we match with the Cardy formula for the string.  

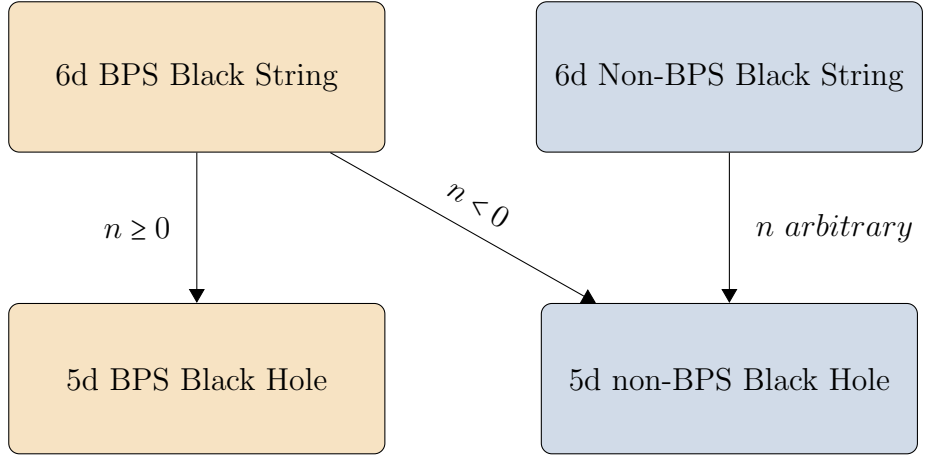
\begin{figure}
\centering
\begin{tikzpicture}[sharp corners=2pt,inner sep=7pt,node distance=.8cm,every text node part/.style={align=center}]

\node[draw, rounded corners, minimum height = 2cm, minimum width = 5cm, fill=mathyellow] (state0){
6d BPS Black String};
\node[draw,below=2cm of state0, rounded corners, minimum height = 2cm, minimum width = 5cm, fill=mathyellow](state2){
5d BPS Black Hole};
\node[draw,right=2cm of state0, rounded corners, minimum height = 2cm, minimum width = 5cm, fill=mathblue](state1){
6d Non-BPS Black String};
\node[draw,below=2cm of state1, rounded corners, minimum height = 2cm, minimum width =5cm, fill=mathblue](state3){
5d non-BPS Black Hole};

\draw[-triangle 60] (state0) -- (state2) node [midway, above, left = 0.1cm]{$n \geq 0$};
\draw[-triangle 60] (state1) -- (state3) node [midway, above, right = 0.1cm]{$n\ arbitrary $};
\draw[-triangle 60] (state0) -- (state3) node [midway, above, rotate = -32]{ $n < 0$};
\end{tikzpicture}
\caption{The connections between the various black objects we consider in 6d and 5d. An arrow represents a dimensional reduction and matching of the central charge in the various pictures. A BPS black string in 6d descends to either a BPS or a non-BPS black hole in 5d, depending on the relative orientation of the momentum along the $S^1$. A non-BPS black string in 6d descends to a non-BPS black hole in 5d.}\label{fig:connections}
\end{figure}

To summarize, we find that the various ways of computing the central charge of the non-BPS strings agree, via the 6d black string calculation from the near-horizon $AdS_3$ space, and from the 5d black hole entropy and matching with the Cardy formula, therefore making a prediction for the central charge of the theory on the non-BPS string. The various connections between solutions are indicated in Fig.~\ref{fig:connections}. 

\subsection{5d black holes}
We now consider the general case of 5d black holes resulting from elliptic CY 3-folds when the wrapping number of the fiber is much larger than the wrapping number of the base, which is the case in which the Cardy formula applies. We will find that the corresponding black holes are either wrapped around (anti)-ample curves in the base manifold, or have $h^{1,1}(B) - 2$ flat directions. We will make the ansatz that the fiber volume $t_0$ shrinks with the wrapping of the fiber $n$, which we will find to be self-consistent. The volume of the elliptically fibered threefold is again
\begin{equation}
\mathcal{V} = \frac{1}{6}\left(C_{000}t_0^3 + 3C_{00\alpha}t_0^2 t^\alpha + 3 t_0 \Omega_{\alpha \beta}t^\alpha t^\beta \right)\, .
\end{equation}
The black hole potential depends on the volume, its derivatives, and second derivatives, and so we can consistently drop the $t_0^3$ term, but cannot make any further simplifications at this point, and so we take
\begin{equation}
\mathcal{V} \simeq \frac{1}{2}\left(C_{00\alpha}t_0^2 t^\alpha + t_0 \Omega_{\alpha \beta}t^\alpha t^\beta \right)\, .
\end{equation}
Taking two derivatives we find the matrix $A_{IJ}$ takes the form
\begin{equation}
  \begin{pmatrix}
A_{00} & A_{0\alpha}\\
A_{\alpha 0 } & A_{\alpha \beta}
\end{pmatrix} = 
 \begin{pmatrix}
C_{00\alpha}t^\alpha & C_{00\alpha}t_0 + \Omega_{\alpha \gamma}t^\gamma\\
C_{00\alpha}t_0 + \Omega_{\alpha \gamma}t^\gamma & t_0 \Omega_{\alpha \beta}
\end{pmatrix} \simeq 
 \begin{pmatrix}
C_{00\alpha}t^\alpha &  \Omega_{\alpha \gamma}t^\gamma\\
 \Omega_{\alpha \gamma}t^\gamma & t_0 \Omega_{\alpha \beta}
\end{pmatrix}\, .
\end{equation}
The inverse matrix in this limit then takes the form
\begin{equation}
A^{-1} = \begin{pmatrix}
-\frac{t_0}{2V_B}&  \frac{t^\alpha}{2V_B}\\
\frac{t^\alpha}{2V_B}&  \frac{1}{t_0}(\Omega^{\alpha \beta} - \frac{t^\alpha t^\beta}{2V_B})
\end{pmatrix}\, ,
\end{equation}
where $\Omega^{\alpha \beta}$ is the inverse of $\Omega_{\alpha \beta}$, and 
\begin{equation}
V_B = \frac{1}{2}\Omega_{\alpha\beta}t^\alpha t^\beta\, .
\end{equation}
Enforcing $\mathcal{V} = 1$, we have $t_0 \simeq 1/V_B$, and so we write
\begin{equation}
A^{-1} = \begin{pmatrix}
-\frac{1}{2V_B^2}&  \frac{t^\alpha}{2V_B}\\
\frac{t^\alpha}{2V_B}&  V_B \Omega^{\alpha \beta} - \frac{t^\alpha t^\beta}{2}
\end{pmatrix}\, ,
\end{equation}
The effective potential is given by
\begin{equation}
V_{eff} = 2\left(-A^{IJ} + \frac{t^I t^J}{2}\right)q_I q_J\, ,
\end{equation}
with the restriction that $\mathcal{V} = 1$.  The fiber only intersects the section, as it has zero intersections with all base divisors, being the inverse image of the self-intersection of an ample divisor in the base. In the limit that the fiber wrapping $n$ is much greater than the wrapping of the base curve, we then have
\begin{equation}
q_0 \simeq n\, , \quad q_\alpha = \Omega_{\alpha\beta}\Sigma^\beta\, ,
\end{equation}
where $\Sigma^\beta$ is the wrapping number of the base curve. The effective potential then takes the form
\begin{align}
V_{eff} &=  2\left(\frac{n^2}{2V_B^2} - \frac{nt^\alpha q_\alpha}{V_B} - V_B \Omega^{\alpha \beta}q_\alpha q_\beta + \frac{(t^\alpha q_\alpha)^2}{2} + \frac{(t^\alpha q_\alpha)^2}{2} + \frac{n^2}{2V_B^2} + \frac{n t^\alpha q_\alpha}{V_B} \right) \nonumber \\
& = 2\left(\frac{n^2}{V_B^2}- V_B \Omega^{\alpha \beta}q_\alpha q_\beta +(t^\alpha q_\alpha)^2 \right)\, .
\end{align}
Differentiating we have
\begin{equation}
\partial_\alpha V_{eff} = 2\left(-\frac{2n^2}{V_B^3}\Omega_\alpha -\Omega_\alpha (C\cdot C) +2 Z_B q_\alpha\right) = 0\, ,
\end{equation}
where $Z_B = t^\alpha q_\alpha$, and $C\cdot C = q_\alpha \Omega^{\alpha \beta} q_\beta$. We then have
\begin{equation}\label{eqn:bothsides}
\Omega_\alpha\left(\frac{2n^2}{V_B^3} + (C \cdot C) \right) = 2q_\alpha Z_B\, .
\end{equation}
Contracting both sides with $t^\alpha$, we find
\begin{equation}
2V_B\left(\frac{2n^2}{V_B^3} + (C \cdot C) \right) = 2Z_B^2\, .
\end{equation}
and so we can conclude
\begin{equation}
\Omega_\alpha = \frac{2V_B q_\alpha}{Z_B}\, ,
\end{equation}
or
\begin{equation}
t^\alpha = \frac{2V_B \Sigma^\alpha}{Z_B}\, ,
\end{equation}
where we have assumed that the multiplicative factors on both sides do not simultaneously vanish, which we will discuss later. Therefore the M2 branes wrap an (anti)-ample curve in the base. 
We can also write
\begin{equation}
V_B = \frac{1}{2}\Omega_{\alpha\beta}t^\alpha t^\beta = 2 \left(\frac{V_B}{Z_B}\right)^2 (C \cdot C)\ , 
\end{equation}
where we note $C\cdot C = \Sigma^\alpha \Omega_{\alpha \beta} \Sigma^\beta$, and so
\begin{equation}
Z_B^2 = 2 V_B (C \cdot C)\, .
\end{equation}
Using this relationship in the above equations we find
\begin{equation}
V_B = \left(\frac{2n^2}{(C \cdot C)}\right)^{1/3}\, .
\end{equation}
Plugging this back into $V_{eff}$ we find
\begin{align}
V_{eff} &= 2\left(n^2\left(\frac{(C \cdot C)}{2n^2}\right)^{2/3} - (C\cdot C)\left(\frac{2n^2}{ (C\cdot C)} \right)^{1/3} + 2(C\cdot C)\left(\frac{2n^2}{(C\cdot C)} \right)^{1/3} \right)\nonumber \\
 & = 2 |n|^{2/3} (C\cdot C)^{2/3} \left(2^{-2/3} + 2^{1/3} \right)\nonumber \\
 & = 3\times 2^{1/3} |n|^{2/3} (C\cdot C)^{2/3}  \, .
\end{align}
Let us now compute the entropy. We have
\begin{align}
S &= 2\pi \left(\frac{1}{6}V_{eff}\right)^{3/4} = 2\pi \left(2^{-2/3} |n|^{2/3}(C\cdot C)^{2/3}\right)^{3/4} = \sqrt{2}\pi \sqrt{|n| (C\cdot C)}\, ,
\end{align}
which agrees with a Cardy formula based on a wrapped string 
\begin{equation}
S_{mirco} = 2\pi \sqrt{\frac{c|n|}{6}}\, ,
\end{equation}
where in the SUSY case (the M2 brane wrapping the fiber and base curve with the same orientation) we should require the string to have $c = c_L =  \simeq 3 C\cdot C$, and in the non-SUSY case (the M2 brane wrapping the fiber and base curve with opposite orientation), we should take $c = c_R \simeq 3 C \cdot C$. Here the $\simeq$ indicates we are working at large charge where the black hole formula should apply, and in this regime we have $c_L = c_R$.

Recall from above that we assumed the scalar factors on either side of Eq.~\ref{eqn:bothsides} did not vanish, which led us to conclude that the M2 brane wrapped an (anti)-ample curve in the base. If we do not make this assumption (i.e. in the case $C\cdot C < 0$), we are led to the conditions that
\begin{equation}
\left(\frac{2n^2}{V_B^3} + (C \cdot C) \right) = 2 Z_B = 0\, .
\end{equation}
This is two equations for $h^{1,1}(B)$ variables, and so these solutions will have $h^{1,1}(B) - 2$ flat directions. Let us calculate the entropy. We have
\begin{equation}
Z_B = 0\, ,
\end{equation}
and
\begin{equation}
V_B = \left(-\frac{2n^2}{C\cdot C} \right)^{1/3}\, .
\end{equation}
Plugging these back into the the effective potential we have
\begin{align}
V_{eff} &= 2\left(n^2 \left(-\frac{C\cdot C}{2n^2} \right)^{2/3} - \left(-\frac{2n^2}{C\cdot C} \right)^{1/3} (C\cdot C)\right)\nonumber \\
& = 2|n|^{2/3}(|C\cdot C|)^{2/3}\left(2^{-2/3} + 2^{1/3} \right)\nonumber \\
& = 3 \times 2^{1/3}|n|^{2/3}(|C\cdot C|)^{2/3}\, ,
\end{align}
which gives the same value of the effective potential as when the M2 branes wrapped an (anti)-ample divisor in the base, but with the sign of $C\cdot C$ flipped.

To summarize the 5d calculation in the large fiber wrapping limit, we find two solutions: the M2 branes either wrap an (anti)-ample curve in the base, or the solution has $Z_B = 0$ with $h^{1,1}(B) - 2$ flat directions, the latter case always being a non-BPS solution. In both cases the wrapping of the fiber can be of either orientation, giving both BPS and non-BPS solutions. In either case the entropy takes the form
\begin{equation}
S = \sqrt{2}\pi \sqrt{n |C \cdot C|}\, ,
\end{equation}
and the expected central charge is then read off as
\begin{equation}
c = 3 |C\cdot C|\, ,
\end{equation}
which agrees with the microscopic calculation in the case of an ample wrapping in the base, regardless of whether the solution is BPS.

\subsection{6d black strings}\label{sec:6dstring}

We now consider the F-theory limit of the elliptically fibered geometry, which gives IIb on the base $B$. By wrapping enough D3 branes around curves in $B$ we get black strings in 6d which are charged under the self-dual two-form gauge fields $B^\alpha$ which number $h^{1,1}(B) - 1$, and then by taking a circle compactification we arrive at the related 5d black holes. Let us then perform the analysis in the 6d theory, following~\cite{Andrianopoli:2007kz}. It was noted in~\cite{Andrianopoli:2007kz} that in 6d there are two types of solutions: BPS solutions, where the D3 branes wrap (anti)-ample curves in $B$,  or solutions where the central charge $Z$ vanishes, in agreement with the solutions that we found in 5d. The effective potential takes the form
\begin{equation}
V_{eff} = \left(-\Omega^{\alpha\beta} + \frac{t^\alpha t^\beta}{\mathcal{V}_B}\right)q_\alpha q_\beta = \left(-(C\cdot C) + \frac{(t^\alpha q_\alpha)^2}{\mathcal{V}_B}\right)) = -(C \cdot C) + Z^2\, ,
\end{equation}
where $Z$ is the central charge
\begin{equation}
Z = \frac{t^\alpha q_\alpha}{\sqrt{\mathcal{V}_B}}\, ,
\end{equation}
and we implicitly fix $\mathcal{V}_B = 1/2$. Differentiating, we find the equations
\begin{equation}
 Z \partial_\alpha Z = 0\, . 
\end{equation}
For the BPS solutions $\partial_\alpha Z = 0$, the equations of motion give
\begin{equation}
t^\alpha = \frac{2 \Sigma^\alpha \mathcal{V}_B}{(t^\lambda q_\lambda)} = \frac{2 \Sigma^\alpha \mathcal{V}_B}{Z_B}\, ,
\end{equation}
in agreement with the 5d black hole calculation, where $Z_B = \sqrt{\mathcal{V}_B} Z$. Again we have
\begin{equation}
Z_B^2 = 2 V_B (C \cdot C)\, ,
\end{equation}
and so the effective potential for these solutions takes the form
\begin{equation}
V_{eff} = \left( -(C\cdot C) + 2(C \cdot C) \right) =  (C\cdot C)\, .  
\end{equation}
The tension of a BPS string is given by the central charge, or square root of the effective potential.

Clearly in the $Z = 0$ solutions, which are only valid for $C \cdot C < 0$, we find 
\begin{equation}
V_{eff} = |C \cdot C|\, ,
\end{equation}
analogous to the BPS solutions, and in agreement with the 5d solutions. Again, such solutions have $h^{1,1}(B) -2$ flat directions. Let us note that not all curves with $C \cdot C \neq 0$ correspond to to large black strings. For instance, if we consider a D3 brane wrapped on a holomorphic curve with negative self-intersection, the non-BPS equations of motion force the horizon moduli to the boundary of the K\"ahler cone where $C$ has zero volume. On the other hand, one can also find examples with $C\cdot C >0$, but $C$ has negative intersection with a holomorphic curve. For example, consider $\mathbb{P}^2$ with projective coordinates $[x_1, x_2, x_3]$, and blow up the locus $x_2 = x_3 = 0$ to get a $dP_1$. The exceptional divisor $D_e$ has self intersection $-1$, the divisor $D_1$ corresponding to $x_1 = 0$ has self intersection $1$, and $D_e \cdot D_1 = 0$. Now consider the curve $D = n D_1 + D_e$, with $n > 1$. We have $D^2 = n^2 - 1$, but $D\cdot D_e = -1$, and so while $D^2 > 0$, $D$ does not correspond to a large black string.

To summarize, the equations of motion for the 5d and 6d cases agree and fix the moduli to the same value, and the wrapping of the fiber should be associated with the momentum of the black string around the circle upon reduction. This is true in both the BPS and non-BPS cases, regardless of whether the associated curve in the base is holomorphic or not. In fact, in both cases the effective potential is given by $V_{eff} = |C \cdot C|$, regardless of whether the string is BPS or not. In \S\ref{central} we show how this leads to the anticipated result for the central charge.

\subsubsection{Measure-theoretic musings}
Let us comment on some measure-theoretic prospects. We note that given a curve class, the effective potential simply produces a number for the volume of the corresponding connected representative, given by the tension
\begin{equation}
T = \sqrt{|C\cdot C|}\, ,
\end{equation}
with the volume of $B$ held fixed to $1/2$, and $Z = 0$. This is somewhat striking, as it predicts that the volume of an LBBC is always given by the absolute value of the self-intersection of the curve, so long as we can solve $\mathcal{V}_B = 1/2$ and $Z = 0$.

We can consider the difference in hyperplane sections $C = C_1 - C_2$ for $B = \mathbb{P}^1 \times \mathbb{P}^1$, as we did in $\S$~\ref{sec:bfn}. In this case the attractor mechanism sets the two K\"ahler moduli (sizes of the $\mathbb{P}^1$'s) to be equal, and we find that the tension of the black string is given by the volume of the minimal piecewise-calibrated representative of $C$. This is different from what the analysis in $\S$~\ref{sec:bfn} suggested, where we found suggestive evidence for recombination. One difference between the two cases is the values of the K\"ahler moduli: in this case the moduli are equal, and in $\S$~\ref{sec:bfn} one modulus was three times the other. 
Indeed the possibility of recombination could depend on the K\"ahler moduli and the discussions in \ref{sec:bfn} was based on the moduli fixed by the attractor mechanism for the string in 5d which is different from that in 6d.  
From the physics perspective this suggests that when we compactify on a circle, the tension of the transverse string (not wrapped around the circle) decreases as we decrease the radius of the circle.  

We can also analyze K3 from the perspective of IIb on K3. To make a connection to the Sen and Micallef-Wolfson examples, let us consider a non-holomorphic curve $C = C_1 - C_2$ of negative self-intersection, where $C_1$ and $C_2$ are both holomorphic. The tension of the black string is then given by
\begin{equation}
T = \sqrt{|C\cdot C|}\, ,
\end{equation}
while the volume of a piecewise-calibrated representative of $C$ is given my
\begin{equation}
\mathrm{vol}^\cup(C) = \mathrm{vol}(C_1) + \mathrm{vol}(C_2)\, .
\end{equation}
In the example of Micallef and Wolfson, $C_1$ and $C_2$ are both $(-2)$-curves of equal volume (normalized to 1), so the black string condition $Z = 0$ is automatically satisfied. This example also has $C\cdot C = -4$, and so in order to find recombination one would then want to tune $\mathrm{vol}^\cup(C)$ to be as large as possible, while holding $\mathrm{vol}(V_{K3})$ fixed. 
Intuitively, recombination in the Micallef-Wolson example occurred when the cycle connecting $C_1$ and $C_2$ shrunk compared to the volumes of $C_1$ and $C_2$, which appears consistent with holding the overall volume fixed while increasing the volumes of $C_1$ and $C_2$. However, in the Micallef-Wolfson example the overall volume of the K3 is taken to be arbitrarily large and essentially decouples from the analysis, which is a local one.

A simpler case than K3 is taking $B$ to be a space with fewer moduli, such as a blowup of $\mathbb{P}^2$ at two distinct points. Let the homogeneous coordinates on $\mathbb{P}^2$ be $[x_1,x_2,x_3]$, and blow up the points $x_1 = x_2 = 0$ and $x_1 = x_3 = 0$. The cone of curves is then generated by $C_1$, $C_a$, and $C_b$, where $C_a$ and $C_b$ are the exceptional divisors from the blowup. Consider the divisor $C = C_a - C_b$, which has self-intersection $(-2)$. Let the volumes of the divisors be $b_1, b_a, b_b$. Enforcing $Z = 0$ gives $b_a = b_b$, and solving $\mathcal{V} = 1/2$ gives
\begin{equation}
b_a = \frac{1}{2}\left(-2 b_1 + \sqrt{2}\sqrt{1 + b_1^2} \right)\, .
\end{equation}
In order to remain in the K\"ahler cone we need $b_1 < 1$. The tension of the non-BPS string is $\sqrt{2}$. In terms of $b_1$, the volume of the piecewise-calibrated representative of $C$ is
\begin{equation}
\mathrm{vol}^\cup (C) = 2 b_a = -2 b_1 + \sqrt{2}\sqrt{1 + b_1^2}\, ,
\end{equation}
which has a maximum of $\sqrt{2}$ at $b_1 = 0$, and shrinks to zero at $b_1 = 1$. Therefore, when we shrink down the cycle connecting $D_a$ and $D_b$ we find that the tension of the non-BPS black string is given by the volume of a piecewise-calibrated representative, but as we move away from this locus the piecewise-calibrated representative has smaller volume, and so we do not find recombination in this example.

\subsubsection{The central charge}\label{central}

To close the circle of ideas, we need to compute the central charge of the 6d string and see if it agrees with that anticipated based on black hole entropy in 5d, which we undertake in this section.
The near-horizon geometry of 6d strings is $AdS_3 \times S^3$, and so associated with and $AdS_3$ solution is a dual 2d conformal field theory. This $CFT_2$ is then expected to describe the worldsheet theory of the associated string, whether it is BPS or not (though in the non-BPS case we expect it to be an ``unstable'' CFT). Let us compute the central charge. In general, via the Brown-Henneaux central charge formula~\cite{cmp/1104114999}, the central charge of the $CFT_2$ is related to the $AdS_3$ radius and Newton's constant
\begin{equation}
c = \frac{3 l_{AdS}}{2 G_3}\, .
\end{equation}
In~\cite{Kraus:2005vz}, a straightforward method was given to compute the central charge, given the attractor values of the scalars. Consider a $d$-dimensional theory specified by a Lagrangian $\mathcal{L}_d$, that admits black object solutions with a near-horizon geometry of the form $AdS_3 \times S^p$. The fields on the horizon are fixed by the attractor mechanism, and the method of~\cite{Kraus:2005vz} is to treat the $AdS_3$ radius $l_{AdS}$ and $p$-sphere radius $l_p$ as free parameters, and extremize a central charge function with respect to $l_{AdS}$ and $l_{S^p}$ to obtain their values, which can roughly be thought of as extremizing the bulk action. One can then evaluate the central charge function to obtain the corresponding central charge. The central charge function can in general be written as
\begin{equation}
c = \frac{3\Omega_2 \Omega_p}{32 \pi G_{p+3}}l_{AdS}^3 l_{S^p}^p \mathcal{L}_{p+3}\, .
\end{equation}

For 6d string in F-theory with have $p = 3$, and the relevant terms\footnote{In this calculation we only consider the leading-order central charge and so do not include corrections to the field strength due to gravitational Chern-Simons terms.} in the action we consider are~\cite{Haghighat:2015ega}
\begin{equation}
\mathcal{S}_6 = \int_{\mathcal{M}_6 }\left[\frac{R}{2} \ast 1-\frac{1}{4}g_{\alpha\beta}H^\alpha \wedge \ast H^\beta-\frac{1}{2}g_{\alpha \beta}dt^\alpha \wedge \ast dt^\beta\right]\, ,
\end{equation}
where the $t^\alpha$ are the K\"ahler moduli on $B$, $g_{\alpha\beta}$ is the metric on moduli space which takes the form
\begin{equation}
g_{\alpha\beta} = \frac{\Omega_\alpha \Omega_\beta}{\mathcal{V}_B} - \Omega_{\alpha\beta}\, ,
\end{equation}
where we implicitly hold $\mathcal{V}_B$ fixed to $1/2$, and $H^\alpha$ are the two-form tensor field strengths $H^\alpha = dB^\alpha$. The integral charges are defined by the fluxes as
\begin{equation}
4 \pi^2 Q^\alpha = \int\limits_{S^3}H^\alpha\, ,
\end{equation}
and so we have
\begin{equation}
H^\alpha = \frac{2}{l_{S^3}^3}Q^\alpha \epsilon_{S^3}\, ,
\end{equation}
where $\epsilon_{S^3}$ is the volume form on the $S^3$. Let us now evaluate the central charge function. We have
\begin{equation}
c = \frac{3\Omega_2 \Omega_p}{32 \pi G_{6}}l_{AdS}^3 l_{S^3}^3 \mathcal{L}_{6}  = \frac{3\pi^2}{4 G_{6}}l_{AdS}^3 l_{S^3}^3 \mathcal{L}_{6} = -\frac{3\pi^2}{4 G_{6}}l_{AdS}^3 l_{S^3}^3\left(-\frac{3}{l_{AdS}^2} + \frac{3}{l_{S^3}^2}  - \frac{g_{\alpha\beta} Q^\alpha Q^\beta}{l_{S^3}^6} \right)\, .
\end{equation}
From the 6d attractor mechanism above, we have that
\begin{equation}
g_{\alpha\beta} Q^\alpha Q^\beta = |C \cdot C|\, ,
\end{equation}
where $C$ is the curve wrapped by the D3 branes. Extremizing the central charge function, we find
\begin{equation}
l_{S^3} = l_{AdS} = \left(\frac{|C \cdot C|}{2} \right)^{1/4}\, ,
\end{equation}
and so the central charge takes the form
\begin{equation}
c = \frac{3\pi^2 |C \cdot C|}{4G_6}\, .
\end{equation}
In these conventions, in order to match the the BPS case we take $G_6 = \frac{\pi^2}{4}$, for which the central charge takes the form
\begin{equation}
c  = 3|C \cdot C|\, .
\end{equation}
However, this formula does not assume the string is BPS, and the results are valid for non-BPS strings as well. Therefore, if a large black string exists for a given curve $C$, the central charge is given by $3|C\cdot C|$, regardless of whether the string is BPS or not. This agrees with what we found in 5d.

\section{Discussion}\label{sec:disc}
In this paper we have studied aspects of non-supersymmetric black holes and black strings in 5d and 6d in theories with 8 supercharges.  This is motivated by the importance of considering non-supersymmetric configurations in string theory.  We have found that extremal non-BPS configurations, at least in the limit of large charges, can have robust features similar to what one sees for supersymmetric BPS states.  Moreover we have explored the interplay between non-BPS black holes and black strings with the Weak Gravity Conjecture, which is motivated by the condition that all non-supersymmetric macroscopic states in string theory are bound to decay.  In a number of examples we have seen that there will have to exist remnant stable non-BPS states, which when combined with the WGC predicts that these strings should be microscopic with small charges. 

Unlike the black strings, for black holes we have found that the non-BPS states seems to decay to BPS and anti-BPS constituents, and we have found no examples of macroscopic black holes whose mass predicts a stable remnant microscopic black hole coming from Calabi-Yau threefolds. However, our microscopic force analysis in \S\ref{sec:bfn} suggests that such a stable remnant might exist, either via recombination or a 5d bound state.
One explanation of this may be the fact that black holes correspond to cycles which are less than half of the dimension of the manifold whereas strings correspond to cycles with dimension bigger than half of the dimension of the manifold.  This would suggest a generic intersection for higher dimensional cycles due to local instability modes localized where holomorphic and anti-holomorphic cycles intersect.

For the case of stable non-BPS strings it would be interesting to confirm the existence of such states mathematically.  This would entail a study of volume minimizing currents in non-holomorphic divisor classes.  This is a rather difficult subject mathematically, and thus the impetus and predictions coming from the work presented here will hopefully lead to further progress in this direction.  Moreover, the fact that we are predicting complex structure-independent minimum volumes for large non-holomorphic classes begs for a mathematical explanation.  This is of course expected for holomorphic ones, but what is the explanation of this behavior for non-holomorphic ones?

The fact that non-BPS extremal strings can account for the entropy of non-BPS black holes is a novel feature that we have found in this paper.  This suggests more broadly that extremal non-supersymmetric configuration, even though ultimately unstable, share features very similar to the supersymmetric counterparts which are stable.  It would be thus interesting to identify such non-supersymmetric extremal configurations either as states or as string compactifications more broadly in hopes of applying them to the observed universe, which is non-supersymmetric.

\section*{Acknowledgments}

We are grateful to Naomi Gendler, Liam McAllister, and John Stout for useful discussions. The
work of C.L. was partially supported by the Alfred P. Sloan Foundation Grant No. G-2019-12504 and by DOE Grant DE-SC0013607. The work of of A. S. was partially supported by the US grants: NSF DMS-1607871, NSF DMS-1306313, Simons 38558, as well as Laboratory of Mirror Symmetry NRU HSE, RF Government
grant, ag. No 14.641.31.0001.   The research of C.V. was partially supported by the
National Science Foundation under Grant No. NSF PHY-2013858 and by a grant from the Simons
Foundation (602883, CV).  The work of S.-T. Y. was partially supported by the US grants: NSF DMS-0804454, NSF PHY1306313, and Simons Foundation 38558.

\bibliographystyle{JHEP}
\bibliography{refs}

\end{document}